\documentclass[aps,twocolumn,twoside,showpacs,preprintnumbers,nofootinbib,superscriptaddress]{revtex4}

\usepackage[dvipdfm,
		colorlinks=true,
		linkcolor=blue,
		urlcolor=magenta,
		bookmarks=true,
		bookmarksnumbered=true,
		bookmarkstype=toc,
		pdftitle={Gauge-Fixing and Residual Symmetries in Gauge/Gravity Theories with Extra Dimensions}]{hyperref}
\usepackage{amsmath}
\usepackage{amssymb}
\usepackage{amsbsy}
\usepackage{mathptmx} 	% MATHPTMX FONT
\usepackage{graphicx}
\usepackage{color}
\usepackage{bm}

\begin{document}
\preprint{KOBE-TH-07-09}
%%%%% TITLE %%%%%
\title{Gauge-Fixing and Residual Symmetries in Gauge/Gravity Theories with Extra Dimensions}
%%%%% AUTHORS %%%%
\author{C. S. Lim}
\email{lim@kobe-u.ac.jp}
\affiliation{Department of Physics, Kobe University, 1-1 Rokkodai, Nada, Kobe 657-8501, Japan}
\author{Tomoaki Nagasawa}
\email{nagasawa@anan-nct.ac.jp}
\affiliation{Anan National College of Technology, 265 Aoki, Minobayashi, Anan 774-0017, Japan}
\author{Satoshi Ohya}
\email{ohya@kobe-u.ac.jp}
\affiliation{Graduate School of Science, Kobe University, 1-1 Rokkodai, Nada, Kobe 657-8501, Japan}
\author{Kazuki Sakamoto}
\email{049d841n@stu.kobe-u.ac.jp}
\affiliation{Graduate School of Science, Kobe University, 1-1 Rokkodai, Nada, Kobe 657-8501, Japan}
\author{Makoto Sakamoto}
\email{dragon@kobe-u.ac.jp}
\affiliation{Department of Physics, Kobe University, 1-1 Rokkodai, Nada, Kobe 657-8501, Japan}
%%%%% DATE %%%%%
\date{\today}
%%%%% ABSTRACT %%%%%
\begin{abstract}
We study compactified pure gauge/gravitational theories with gauge-fixing terms and show that these theories possess quantum mechanical SUSY-like symmetries between unphysical degrees of freedom.
These residual symmetries are global symmetries and generated by quantum mechanical ${\cal N}=2$ supercharges.
Also, we establish new one-parameter family of gauge choices for higher-dimensional gravity, and calculate as a check of its validity one graviton exchange amplitude in the lowest tree-level approximation.
We confirm that the result is indeed $\xi$-independent and the cancellation of the $\xi$-dependence is ensured by the residual symmetries.
We also give a simple interpretation of the vDVZ-discontinuity, which arises in the lowest tree-level approximation, from the supersymmetric point of view.
\end{abstract}
%%%%% PACS CODE %%%%%
\pacs{11.30.Pb, 11.10.Kk}
\maketitle
%%%%% SECTION 1 %%%%%
\section{Introduction} \label{sec:intro}
One of the salient features in compactified higher-dimensional field theories is the geometric ``Higgs'' mechanism resulting from the dimensional reduction.
If gauge particles can propagate to extra dimensions, the corresponding higher-dimensional gauge fields should be decomposed into normal modes of extra dimensions and then, from the four-dimensional point of view, they can be recast into an infinite tower of massive gauge bosons whose longitudinal degree of freedom is provided by absorbing one of extra spatial components of the field.
This gauge boson mass generation is realized without invoking any scalar Higgs fields; it is essentially geometrical in nature.
It also occurs in compactified higher-dimensional gravity theories the same mass generation mechanism without invoking any fundamental Higgs fields.
There, it seems that there is no explicit symmetry breaking.
However, it was shown in the Kaluza-Klein (KK) theory that the spontaneous symmetry breaking certainly occurs \cite{DD:1984,Aulakh:1985un,Witten:1981me,Salam:1981xd,CZ:1992}: there exists an infinite-dimensional Kac-Moody-like symmetry at the 4d Lagrangian level, but it is broken down to the 4d translation and internal U(1) symmetry by the vacuum configuration $M^{4}\times S^{1}$, and then the non-zero graviton modes become massive.
Though it would be the best to describe the geometric ``Higgs'' mechanism by the argument along this line, it seems to be hard to extend the analysis to other more complicated compactified gravity theories.
It is less obvious such analysis is applicable to compactified gauge theories.

Recently, an alternative view of this mass generation mechanism is given by \cite{LNSS:2005,LNOSS1:2007}: it is best described by using the hidden quantum mechanical supersymmetry in 4d mass spectrum.
The point is that mass eigenfunction for each non-zero KK-mode has its own superpartner.
These functions are defined as eigenfunctions of $2\times2$ matrix super-Hamiltonian.
Therefore it can be explained that $n$th KK-mode of the 4d component of the gauge field can absorb as its longitudinal degrees of freedom its partner $n$th KK-mode of the extra component of the field, and then $n$th gauge boson becomes massive.

In ordinary spontaneously broken gauge theories such as the Abelian Higgs model, if we work in the $R_{\xi}$ gauge it is inevitable to introduce the fictitious particles, i.e. the would-be Nambu-Goldstone (NG) bosons, so as to maintain the unitarity of the $S$-matrix.
The scalar component of the massive gauge field, the would-be NG boson, the Faddeev-Popov (FP) ghost and the anti FP ghost are all in degenerate with the same mass-squared $\xi m^{2}$.
These spurious degrees of freedom consist of a multiplet, known as the BRS quartet \cite{KO:1979}, and do not contribute to the physical amplitude by canceling each other, and then the unitarity of the $S$-matrix is maintained.

However, there is no such scalar Higgs fields in pure gauge theory with extra compact dimensions, nor in compactified higher-dimensional pure gravity.
Thus the unphysical polarization states of higher-dimensional gauge/gravitational fields must be canceled among themselves.
Furthermore, in order to realize this cancellation these spurious degrees of freedom must be in degenerate with the same gauge-dependent mass-squared.
In view of this, we have to guarantee at least two things: the first is the same number of degrees of freedom for the unphysical components of 4d gauge/gravitational field and the would-be NG bosons.
These degrees of freedom have to appear in pairs.
The second is the degeneracy of mass, that is, the same $\xi$-dependent pole of the propagators between these pairs.
In view of this, it is sufficient to consider pure gauge/gravitational theories up to quadratic order and compute the propagators in $R_{\xi}$ gauges.
It should be noted at this stage that, from the knowledge of the supersymmetric structure in the 4d mass spectrum, the mass matrices of these pairs should be given by the ${\cal N} = 2$ super-Hamiltonians.
Therefore it is reasonable to expect that these free field theories would be invariant under some SUSY-like transformations generated by supercharges to rotate these unphysical degrees of freedom with different spins.
This is the main subject of this paper.
We study the compactified pure Abelian gauge theory and pure gravity to quadratic order of gravitational fluctuations.
We show that these free field theories exhibit residual global SUSY-like symmetries between the unphysical components of 4d gauge/gravitational fields and the would-be Nambu-Goldstone bosons, which are one of the extra space components of the higher-dimensional fields.
Also, we establish new $R_{\xi}$ gauges in five-dimensional gravity.
We check its validity by computing the lowest tree-level graviton exchange amplitude.
Of course the result does not depend on the gauge parameter $\xi$.
Note that we will not argue with unitarity bounds of some specific models as discussed in \cite{CDH:2002,CDHN:2003}.
In this paper we will restrict ourselves to free field theories.

The rest of this paper is organized as follows.
In Section \ref{sec:Abelian} we study the pure Abelian gauge theory with extra $D$-dimensions compactified on a Riemannian manifold and show that in the $R_{\xi}$ gauge the scalar component of the four-dimensional gauge field and the would-be scalar NG boson are in degenerate with the same gauge-dependent mass-squared, and they form a multiplet under the SUSY-like transformation generated by the supercharge found in the analysis of the 4d mass spectrum.
In Section \ref{sec:gravity} we extend the analysis to five-dimensional gravity with the Randall-Sundrum background and obtain the similar results.
In this section we establish the one-parameter family of gauge choices analogous to the $R_{\xi}$ gauge in spontaneously broken gauge theories.
As a check of validity of this $R_{\xi}$ gauge, we compute the one graviton exchange amplitude in the lowest tree-level and show that the result is indeed $\xi$-independent.
From the result obtained by this computation, we give a simple interpretation of the appearance of the van Dam-Veltman-Zakharov (vDVZ) discontinuity from supersymmetric viewpoint.
Since it seems to be unfamiliar to particle theorists, we devote Appendix \ref{appendix:vDVZ} to a brief review of the vDVZ-discontinuity.
We conclude in Section \ref{sec:summary}.

%%%%% SECTION 2 %%%%%
\section{Pure Abelian Gauge Theory with Extra Dimensions} \label{sec:Abelian}
In this section we show that any compactified pure Abelin gauge theories intrinsically possess hidden quantum mechanical SUSY-like symmetry between a scalar component of 4d gauge field and one of scalar components of higher-dimensional gauge field, which should be regarded as a would-be scalar NG boson.

Let us consider the pure Abelian gauge theory with extra $D$-dimensions compactified on a Riemannian manifold $K$ without boundary.
We denote the $(4+D)$-dimensional gauge field by $A_{M}(x,y) = \bigl(A_{\mu}(x,y), A_{i}(x,y)\bigr)$ $(\mu = 0,1,2,3; i = 5,\cdots, D+4)$, $x^{\mu}$ are the coordinates of the ordinary four-dimensional Minkowski spacetime $M^{4}$ and $y^{i}$ are the coordinates of $K$.
The bulk metric is given by
\begin{equation}
{\rm d}s^{2}
= 	G_{MN}{\rm d}x^{M}{\rm d}x^{N}
= 	\eta_{\mu\nu}{\rm d}x^{\mu}{\rm d}x^{\nu}
	+ g_{ij}(y){\rm d}y^{i}{\rm d}y^{j}, \label{eq:Abelian01}
\end{equation}
where $\eta_{\mu\nu}$ is the flat spacetime metric whose signature is $(-,+,+,+)$.

The action we consider is
\begin{equation}
S
= 	\int_{M^{4}}\!\!\!\!{\rm d}^{4}x\int_{K}\!\!{\rm d}^{D}y\sqrt{-G}
	\left\{-\frac{1}{4}G^{MK}G^{NL}F_{MN}F_{KL}\right\}, \label{eq:Abelian02}
\end{equation}
where $F_{MN} = \partial_{M}A_{N} - \partial_{N}A_{M}$ and $G = {\rm det}(G_{MN}) (= -g)$.
The action is invariant under the $U(1)$ gauge transformation
\begin{equation}
A_{M}(x,y) \mapsto A_{M}^{'}(x,y) = A_{M}(x,y) + \partial_{M}\epsilon(x,y), \label{eq:Abelian03}
\end{equation}
where $\epsilon$ is an arbitrary function.
For the following discussion it is highly convenient to use the elegant mathematical tool of the differential forms.
Notice that on the $D$-dimensional manifold $K$, $A_{\mu}$ and $A_{i}$ behave as a scalar and vector fields such that they can be regarded as 0-forms and a 1-form on $K$, respectively.
For further discussion it is also convenient to introduce the inner product of differential forms on the manifold $K$.
The inner product of two $k$-forms on $K$ is defined by
\begin{equation}
\bigl(\omega^{(k)}, \eta^{(k)}\bigr)
:= 	\int_{K}\omega^{(k)}\wedge*\eta^{(k)}
= 	\int_{K}\!\!{\rm d}^{D}y\sqrt{g}~\frac{1}{k!}\omega^{(k)}_{I}g^{IJ}\eta^{(k)}_{J}, \label{eq:Abelian04}
\end{equation}
where $*$ is the Hodge star operator and
\begin{subequations}
\begin{align}
\omega^{(k)}
&:= 	\frac{1}{k!}\omega^{(k)}_{i_{1}\cdots i_{k}}{\rm d}y^{i_{1}}\wedge\cdots\wedge{\rm d}y^{i_{k}}
=: 	\frac{1}{k!}\omega^{(k)}_{I}{\rm d}y^{I}, \label{eq:Abelian05a}\\
g^{IJ}
&:= 	g^{i_{1}j_{1}}\cdots g^{i_{k}j_{k}}. \label{eq:Abelian05b}
\end{align}
\end{subequations}
$I$ and $J$ denote k-tuples of indices.
With this definition, \eqref{eq:Abelian02} and \eqref{eq:Abelian03} can be written as the following form:
\begin{subequations}
\begin{align}
S
&= 	\int_{M^{4}}\!\!\!\!{\rm d}^{4}x~{\cal L}_{K}, \label{eq:Abelian06a}\\
{\cal L}_{K}
&= 	-\frac{1}{4}\bigl(F_{\mu\nu}, F^{\mu\nu}\bigr)
	- \frac{1}{2}\bigl(\partial_{\mu}A^{(1)} - {\rm d}A_{\mu},
		\partial^{\mu}A^{(1)} - {\rm d}A^{\mu}\bigr) \nonumber\\
&	\hspace{1em}
	- \frac{1}{2}\bigl({\rm d}A^{(1)}, {\rm d}A^{(1)}\bigr), \label{eq:Abelian06b}
\end{align}
\end{subequations}
and
\begin{subequations}
\begin{align}
A_{\mu}
&\mapsto A_{\mu}^{\prime} = A_{\mu} + \partial_{\mu}\epsilon, \label{eq:Abelian07a}\\
A^{(1)}
&\mapsto A^{(1)\prime} = A^{(1)} + {\rm d}\epsilon, \label{eq:Abelian07b}
\end{align}
\end{subequations}
where $A^{(1)} = A_{i}{\rm d}y^{i}$ and ${\rm d} = {\rm d}y^{i}\partial_{i}$.

To clarify the supersymmetric structure, we will follow the method of ref.\cite{LNSS:2005}.
The Hodge decomposition theorem tells us that any $k$-form on $K$ can be {\rm uniquely} decomposed into the sum of the harmonic $k$-form, the exact $k$-form and the coexact $k$-form on $K$.
Noting that there is no exact 0-form, we can write
\begin{subequations}
\begin{align}
A^{(0)}_{\mu}
&= 	A_{\mu,0}(x)\eta^{(0)}
	+ \sum_{n\neq0}A_{\mu,n}(x)\omega^{(0)}_{n}, \label{eq:Abelian08a}\\
A^{(1)}
&= 	\varphi + h + \Phi \nonumber\\
&= 	\sum_{i=1}^{b_{1}}\varphi_{i}(x)\eta^{(1)}_{i}
	+ \sum_{n\neq0}h_{n}(x)\phi^{(1)}_{n}
	+ \sum_{n^{'}\neq0}\Phi_{n^{'}}(x)\omega^{(1)}_{n^{'}}, \label{eq:Abelian08b}
\end{align}
\end{subequations}
where $\eta^{(0)} = \left[\int_{K}{\rm d}^{D}y\sqrt{g}\right]^{-1/2}$ and $b_{1}$ is the first Betti number.
As we shall see immediately, $b_{1}$ gives the number of 4d massless scalar bosons, which are identified as Higgs fields in the gauge-Higgs unification scenario.
For example, $b_{1} = 1$ for $K = S^{1}$, $b_{1} = 0$ for $K = S^{2}$, $b_{1} = \binom{n}{1} = n$ for $K = T^{n} = S^{1}\times\cdots\times S^{1}$ (K\"unneth formula), etc.

$\omega^{(k)}_{n}$ and $\phi^{(k+1)}_{n}$ ($k=0,1$) are coexact $k$-form and exact $k+1$-form and defined as the orthonormal sets of the eigenmodes of the positive semi-definite isospectral Hamiltonians ${\rm d}^{\dagger}{\rm d}$ and ${\rm d}{\rm d}^{\dagger}$:
\begin{equation}
\begin{bmatrix}
{\rm d}^{\dagger}{\rm d} 	& 0 \\
0 						& {\rm d}{\rm d}^{\dagger}
\end{bmatrix}
\begin{bmatrix}
\omega^{(k)}_{n} \\
\phi^{(k+1)}_{n}
\end{bmatrix}
= \bigl(m^{(k)}_{n}\bigr)^{2}
\begin{bmatrix}
\omega^{(k)}_{n} \\
\phi^{(k+1)}_{n}
\end{bmatrix}, \label{eq:Abelian09}
\end{equation}
where the relative phase is chosen to be
\begin{equation}
\begin{bmatrix}
0 		& {\rm d}^{\dagger} \\
{\rm d} 	& 0
\end{bmatrix}
\begin{bmatrix}
\omega^{(k)}_{n} \\
\phi^{(k+1)}_{n}
\end{bmatrix}
= m^{(k)}_{n}
\begin{bmatrix}
\omega^{(k)}_{n} \\
\phi^{(k+1)}_{n}
\end{bmatrix}. \label{eq:Abelian10}
\end{equation}
Now we introduce the following $2\times2$ matrix operators
\begin{equation}
H = 	\begin{bmatrix}
	{\rm d}^{\dagger}{\rm d} 	& 0 \\
	0 						& {\rm d}{\rm d}^{\dagger}
	\end{bmatrix}, \quad
Q_{1} = 	\begin{bmatrix}
	0 		& {\rm d}^{\dagger} \\
	{\rm d} 	& 0
	\end{bmatrix}, \quad
(-1)^{F} = 	\begin{bmatrix}
		1 	& 0 \\
		0 	& -1
		\end{bmatrix}, \label{eq:Abelian11}
\end{equation}
which satisfy the ${\cal N} = 2$ supersymmetry algebra
\begin{align}
& 	\{Q_{a}, Q_{b}\} = 2\delta_{ab}H, \quad
	[H, Q_{a}] = [H, (-1)^{F}] = 0, \nonumber\\
& 	\{(-1)^{F}, Q_{a}\} = 0, \quad
	a,b = 1,2, \label{eq:Abelian12}
\end{align}
where $Q_{2} = i(-1)^{F}Q_{1}$.
$(-1)^{F}$ is the ``fermionic'' number operator which assigns to the ``bosonic'' state $\omega^{(k)}_{n}$ a quantum number $+1$ and to the ``fermionic'' state $\phi^{(k)}_{n}$ a quantum number $-1$.

From the orthonormality of the eigenmodes
\footnote{
The eigenmodes are normalized as
\begin{align}
& 	\bigl(\omega^{(k)}_{n}, \omega^{(k)}_{m}\bigr) = \delta_{nm}, \quad
	\bigl(\phi^{(k+1)}_{n}, \phi^{(k+1)}_{m}\bigr) = \delta_{nm}, \nonumber\\
& 	\bigl(\eta^{(1)}_{i}, \eta^{(1)}_{j}\bigr) = \delta_{ij}. \nonumber
\end{align}}
, it is easy to write down the 4d reduced Lagrangian
\begin{align}
{\cal L}_{K}
&= 	- \frac{1}{4}\bigl(F_{\mu\nu,0}\bigr)^{2} \nonumber\\
&	+ \sum_{n\neq0}
	\left\{-\frac{1}{4}\bigl(F_{\mu\nu,n}\bigr)^{2}
	- \frac{1}{2}\bigl(m_{n}^{(0)}\bigr)^{2}
	\left(A_{\mu,n} - \frac{1}{m_{n}^{(0)}}\partial_{\mu}h_{n}\right)^{2}\right\} \nonumber\\
&	+ \sum_{i=1}^{b_{1}}
	\left\{-\frac{1}{2}\bigl(\partial_{\mu}\varphi_{i}\bigr)^{2}\right\} \nonumber\\
&	+ \sum_{n^{'}\neq0}
	\left\{- \frac{1}{2}\bigl(\partial_{\mu}\Phi_{n^{'}}\bigr)^{2}
	- \frac{1}{2}\bigl(m_{n^{'}}^{(1)}\bigr)^{2}\Phi_{n^{'}}^{2}\right\}, \label{eq:Abelian13}
\end{align}
where $F_{\mu\nu,n} = \partial_{\mu}A_{\nu,n} - \partial_{\nu}A_{\mu,n}$.
It is worth mentioning at this point that each KK-mode transforms under the U(1) gauge transformation \eqref{eq:Abelian07a} \eqref{eq:Abelian07b} as $\delta A_{\mu,0} = \partial_{\mu}\epsilon_{0}$, $\delta A_{\mu,n} = \partial_{\mu}\epsilon_{n}$, $\delta h_{n} = m^{(0)}_{n}\epsilon_{n}$ and the others are equal to zero.
Hence, from the higher-dimensional gauge invariant point of view, $h_{n}$ has to appear to ${\cal L}_{K}$ in the combination $A_{\mu,n} - (1/m^{(0)}_{n})\partial_{\mu}h_{n}$, which is realized in \eqref{eq:Abelian13} as it should be.
By (partially) fixing the gauge $\epsilon_{n} = - (1/m^{(0)}_{n})h_{n}$ (with $\epsilon_{0}$ left arbitrary) all the would-be NG bosons $h_{n}$ are gauged away.
This is the unitary gauge of the theory.
The four-dimensional particle content of the theory is now obvious: the massless photon $A_{\mu,0}$, the infinite tower of massive gauge bosons $A_{\mu,n}$ with mass $m_{n}^{(0)}$, the $b_{1}$ massless scalar bosons $\phi_{i}$ and the infinite tower of massive scalar bosons $\Phi_{n^{'}}$ with mass $m_{n^{'}}^{(1)}$.

%%%%% SUB-SECTION 2.1 %%%%%
\subsection{Gauge-Fixing and Residual Symmetry} \label{subsec:Abelian_gauge-fix}
As already mentioned in Section \ref{sec:intro}, in this paper we are interested in the $R_{\xi}$ gauge.
Since there is an undesirable quadratic term mixing $A_{\mu}$ and $A^{(1)}$ in \eqref{eq:Abelian06b} (or $A_{\mu,n}$ and $h_{n}$ in \eqref{eq:Abelian13}), we have to get rid of this mixing term by correctly choosing a gauge-fixing function.
Such a gauge-fixing function is already introduced, for example, in \cite{MPR:2002} in the context of five-dimensional gauge theory.
The $D$-dimensional version of such a gauge-fixing function should be chosen as
\begin{equation}
F[A]
= 	\partial^{\mu}A_{\mu} - \xi{\rm d}^{\dagger}A^{(1)}
= 	\partial^{\mu}A_{\mu} - \xi{\rm d}^{\dagger}h. \label{eq:Abelian14}
\end{equation}
By adding a gauge-fixing term $-(1/2\xi)F[A]^{2}$ to \eqref{eq:Abelian06b} and integrating by parts we obtain the gauge-fixed Lagrangian
\begin{align}
{\cal L}
&= 	\frac{1}{2}\bigl(A_{\mu}, [\eta^{\mu\nu}(\Box - {\rm d}^{\dagger}{\rm d})
		- (1 - 1/\xi)\partial^{\mu}\partial^{\nu}]A_{\nu}\bigr) \nonumber\\
&	\hspace{1em}
	+ \frac{1}{2}\bigl(\varphi, \Box\varphi\bigr)
	+ \frac{1}{2}\bigl(\Phi, [\Box - {\rm d}^{\dagger}{\rm d}]\Phi\bigr) \nonumber\\
& 	\hspace{1em}
	+ \frac{1}{2}\bigl(h, [\Box - \xi{\rm d}{\rm d}^{\dagger}]h\bigr), \label{eq:Abelian15}
\end{align}
where $\Box = \partial_{\mu}\partial^{\mu}$.
Inverting the kinetic term operator we obtain the gauge boson $A_{\mu,n}$ propagator containing a term proportional to $k_{\mu}k_{\nu}/\bigl(m_{n}^{(0)}\bigr)^{2}$ with the gauge-dependent pole at $- k^{2} = \xi\bigl(m_{n}^{(0)}\bigr)^{2}$, which is exactly the same simple pole position of the would-be NG boson's ($h_{n}$'s), as it should.
In any free field theories when two different fields are in degenerate and have the same degrees of freedom, there obviously exists a symmetry to rotate them.
Now this is indeed the case.
The term proportional to $k_{\mu}k_{\nu}/\bigl(m_{n}^{(0)}\bigr)^{2}$ with the gauge-dependent pole at $- k^{2} = \xi\bigl(m_{n}^{(0)}\bigr)^{2}$ in the gauge boson propagator suggests that the scalar component of the gauge field has the same mass-squared $\xi \bigl(m_{n}^{(0)}\bigr)^{2}$ to the would-be NG boson's.
Indeed, if we decompose $A_{\mu}$ into the transverse and longitudinal
\footnote{We here use the terms transverse and longitudinal in a covariant way, that is, transverse means perpendicular to $\partial_{\mu}$ (or $k_{\mu}$) and longitudinal parallel to $\partial_{\mu}$ (or $k_{\mu}$).}
 (or pure gauge) part
\begin{equation}
A_{\mu} = A_{\mu}^{({\rm T})} + \partial_{\mu}A^{({\rm L})}, \label{eq:Abelian16}
\end{equation}
the Lagrangian becomes
\begin{align}
{\cal L}
&= 	\frac{1}{2}\bigl(A^{({\rm T})}_{\mu}, [\Box - {\rm d}^{\dagger}{\rm d})]A^{({\rm T})\mu}\bigr)
	- \frac{1}{2\xi}\bigl(A^{({\rm L})}, \Box[\Box - \xi{\rm d}^{\dagger}{\rm d}]A^{({\rm L})}\bigr) \nonumber\\
&	\hspace{1em}
	+ \frac{1}{2}\bigl(\varphi, \Box\varphi\bigr)
	+ \frac{1}{2}\bigl(\Phi, [\Box - {\rm d}^{\dagger}{\rm d}]\Phi\bigr) \nonumber\\
& 	\hspace{1em}
	+ \frac{1}{2}\bigl(h, [\Box - \xi{\rm d}{\rm d}^{\dagger}]h\bigr). \label{eq:Abelian17}
\end{align}
To canonicalize the Lagrangian we redefine the field as
\footnote
{%
The definition \eqref{eq:Abelian18} is well-defined only in the region inside (outside) the light-cone and $\xi>0$ ($\xi<0$.)}
\begin{equation}
A^{({\rm L})} \mapsto {\tilde A}^{({\rm L})} := \sqrt{\frac{\Box}{\xi}}A^{({\rm L})}, \label{eq:Abelian18}
\end{equation}
so that $({\tilde A}^{(L)}, h)$-sector can then be recast into the following $2\times2$ matrix form
\begin{align}
	\frac{1}{2}
		\left(\raisebox{2.5ex}{$t$}\!
		\begin{bmatrix}
		{\tilde A}^{({\rm L})} \\
		h
		\end{bmatrix},
		\begin{bmatrix}
		-1 	& 0 \\
		0 	& 1
		\end{bmatrix}
		\left(\Box - \xi
		\begin{bmatrix}
		{\rm d}^{\dagger}{\rm d} 	& 0 \\
		0 						& {\rm d}{\rm d}^{\dagger}
		\end{bmatrix}\right)
		\begin{bmatrix}
		{\tilde A}^{({\rm L})} \\
		h
		\end{bmatrix}\right). \label{eq:Abelian19}
\end{align}
We have thus separated the Lagrangian into the gauge-independent and -dependent parts.
The gauge-dependent part can be written in terms of $2\times2$ matrix form and the mass matrix is exactly the Hamiltonian in \eqref{eq:Abelian11}, as expected.
It is now obvious the Lagrangian is invariant under the transformation
\begin{equation}
\begin{bmatrix}
{\tilde A}^{({\rm L})} \\
h
\end{bmatrix}
\mapsto
\begin{bmatrix}
{\tilde A}^{({\rm L})\prime} \\
h^{\prime}
\end{bmatrix}
= \theta
\begin{bmatrix}
0 		& {\rm d}^{\dagger} \\
{\rm d} 	& 0
\end{bmatrix}
\begin{bmatrix}
{\tilde A}^{({\rm L})} \\
h
\end{bmatrix}. \label{eq:Abelian20}
\end{equation}
This symmetry transformation is generated by the supercharge $Q_{1}$ in \eqref{eq:Abelian11}, and hence SUSY-like.

Several comments are now in order:

(1) This is a symmetry between the scalar component of the gauge field $A_{\mu}$ and the would-be NG boson $h$, both are unphysical degrees of freedom of the theory.
Thus if we add to the Lagrangian the gauge invariant interaction terms or extend to non-Abelian gauge theories, and then calculate the $S$-matrix elements, these spurious degrees of freedom must be canceled so as to maintain the unitarity of the $S$-matrix.
We may therefore conclude that these cancellations are compensated for the symmetry \eqref{eq:Abelian20}, though it is much less obvious whether residual symmetry such as \eqref{eq:Abelian20} might exist in a fully interacting theory.

(2) As already mentioned in Section \ref{sec:intro}, \eqref{eq:Abelian20} is an expected symmetry.
From the unitarity point of view, these unphysical gauge-dependent part must have its partners to ensure the cancellation of each contributions to the $S$-matrix elements.
In view of this, these partners must be in degenerate with the same $\xi$-dependent mass-squared.
Since the super-Hamiltonian $H$ is only the candidate for the mass-matrix and $H$ is written by product of the supercharges, it is quite natural these spurious degrees of freedom are mutually related by the global transformation generated by $Q_{1}$.

(3) Since the residual symmetry \eqref{eq:Abelian20} is a spacetime symmetry between a spin-1 and a spin-0 particle, one might suspect that \eqref{eq:Abelian20} would conflict with the Coleman-Mandula theorem \cite{CM:1967}.
However, the residual symmetry rotates only the unphysical degrees of freedom.
Hence this is not the case.

(4) This residual symmetry can be regarded as a relic of higher-dimensional gauge symmetry by using the gauge-fixing condition $F[A] = 0$.
It is easy to see that the transformation \eqref{eq:Abelian20} is indeed the higher-dimensional gauge transformation with an identification $\epsilon = \theta{\rm d}^{\dagger}h = \theta(1/\xi)\partial^{\mu}A_{\mu} = \theta(\Box/\xi)A^{({\rm L})}$.
In view of this fact, one might expect that compactified gauge theories with gauge invariant interaction terms would also possess some residual symmetries such as \eqref{eq:Abelian20}.
However, this is an open question.

%%%%% SECTION 3 %%%%%
\section{5d Gravity with Randall-Sundrum Background} \label{sec:gravity}
We would like to further extend the previous analysis to compactified higher-dimensional pure gravity to quadratic order of gravitational fluctuations.
Because of authors' inability to analyze the gravity in general dimensions with general background metric, we shall restrict ourselves to consider the perturbation to the well-known Randall-Sundrum metric \cite{RS1:1999} as a less trivial example.

Let us consider the five-dimensional gravity with single extra dimension compactified on an interval $(0, \pi R)$ with two 3-branes which are located at the end points of the interval.
The action respecting this configuration is
\begin{align}
S
&= 	\int{\rm d}^{4}x\int_{0}^{\pi R}\!\!\!\!{\rm d}y
	\biggl\{\sqrt{-G}(M^{3}R - \Lambda)
	+ \sqrt{-g_{\rm UV}}(-\sigma_{\rm UV})\delta (y) \nonumber\\
&	\hspace{7em}
	+ \sqrt{-g_{\rm IR}}(-\sigma_{\rm IR})\delta (y - \pi R)\biggr\}, \label{eq:gravity01}
\end{align}
where $M$ is the five-dimensional Planck mass, $R$ is the Ricci scalar
\footnote
{%
In this paper we use the conventions:
\begin{align}
\Gamma^{A}_{MN}
&:= 	\frac{1}{2}G^{AB}(\partial_{N}G_{BM} + \partial_{M}G_{BN} - \partial_{B}G_{MN}), \nonumber\\
{R^{K}}_{LMN}
&:= 	\partial_{M}\Gamma^{K}_{LN} - \partial_{N}\Gamma^{K}_{LM}
	+ \Gamma^{A}_{LN}\Gamma^{K}_{AM} - \Gamma^{A}_{LM}\Gamma^{K}_{NA}, \nonumber\\
R_{MN}
&:= 	{R^{A}}_{MAN}
= 	\partial_{A}\Gamma^{A}_{MN} - \partial_{N}\Gamma^{A}_{MA}
	+ \Gamma^{A}_{MN}\Gamma^{B}_{AB} - \Gamma^{B}_{MA}\Gamma^{A}_{NB}. \nonumber
\end{align}
}
evaluated by the 5d metric $G_{MN}$ and $g^{\rm UV}_{\mu\nu}$ ($g^{\rm IR}_{\mu\nu}$) is the metric induced on the UV (IR) brane.
The bulk cosmological constant $\Lambda$ and the brane cosmological constants $\sigma_{\rm UV}$ and $\sigma_{\rm  IR}$ are tuned to give a warped background solution with a slice of ${\rm AdS}_{5}$:
\begin{equation}
{\rm d}s^{2} = 	\begin{cases}
			{\rm e}^{-2ky}\eta_{\mu\nu}{\rm d}x^{\mu}{\rm d}x^{\nu} + {\rm d}y^{2},
			& \textrm{(physical coordinate)} \\
			\left(\frac{1}{1 + kz}\right)^{2}(\eta_{\mu\nu}{\rm d}x^{\mu}{\rm d}x^{\nu} + {\rm d}z^{2}),
			& \textrm{(conformal coordinate)}
			\end{cases} \label{eq:gravity02}
\end{equation}
with $\sigma_{\rm UV} = 12kM^{3}$ and $\sigma_{\rm IR} = -12kM^{3}$, where $k$ is the AdS curvature scale defined as $k = \sqrt{(-\Lambda)/12M^{3}}$.
It is just a matter of choice which coordinates (physical or conformal) one will use, however, the conformal coordinate will turn out to be advantageous to evaluate various quantities, especially the gravity action because it utilizes the very useful formula of the conformal transformation.
Thus in what follows we will use the conformal coordinate $z$ for computational convenience.

We investigate the gravitational fluctuations around the background
\begin{align}
G_{MN}
= 	{\rm e}^{2A(z)}\bigl[\eta_{MN} + {\bar h}_{MN}(x)\bigr], \quad
A(z)
= -\ln(1 + kz). \label{eq:gravity03}
\end{align}
It should be noted that the explicit expression for the warp factor $A(z)$ is valid on the {\em open} interval $0 < z < z_{c} := (1/k)({\rm e}^{\pi kR} - 1)$:
\begin{align}
A'' - (A')^{2} = 2k{\rm e}^{A}[- \delta(z) + \delta(z - \pi R)], \label{eq:gravity04}
\end{align}
where prime (${}^{\prime}$) indicates the derivative with respect to $z$.
In the following analysis, however, the delta functions on the right hand side of Eq.\eqref{eq:gravity04} can safely be ignored because of the boundary conditions for the metric fluctuations consistent with the general coordinate invariance.
Therefore we will use the relation $A^{''} - (A^{'})^{2} = 0$ in the whole of this paper instead of \eqref{eq:gravity04}.

The fluctuations ${\bar h}_{MN}$ should be parameterized as
\begin{equation}
{\bar h}_{MN}
= 	\begin{bmatrix}
	h_{\mu\nu} - (1/2)\eta_{\mu\nu}\phi 	& h_{\mu5} \\
	h_{\nu5} 							& \phi
	\end{bmatrix}, \label{eq:gravity05}
\end{equation}
which enables us to identify the spectrum of the linearized theory.
Notice that the limit $\Lambda \to 0$ gives $k=0$ and $z_{c} = \pi R$ and hence the background geometry \eqref{eq:gravity02} just reduces to $M^{4}\times (0,\pi R)$, as it should.

The quadratic action takes the form
\begin{align}
S^{(2)}
&= 	M^{3}\!\!\int{\rm d}^{4}x\int_{0}^{z_{c}}\!\!\!{\rm d}z~{\rm e}^{3A}
	\frac{1}{4}h^{MN}K_{MN; KL}h^{KL} \nonumber\\
&= 	M^{3}\!\!\int{\rm d}^{4}x\int_{0}^{z_{c}}\!\!\!{\rm d}z~{\rm e}^{3A} \nonumber\\
&	\hspace{1em}
	\times
	\frac{1}{4}
	\biggl\{h^{\mu\nu}K_{\mu\nu; \rho\sigma}h^{\rho\sigma}
	+ 2h^{\mu\nu}K_{\mu\nu; \rho5}h^{\rho5}
	+ h^{\mu\nu}K_{\mu\nu; \phi}\phi \nonumber\\
& 	\hspace{2em}
	+ 2h^{\mu5}K_{\mu5; \rho\sigma}h^{\rho\sigma}
	+ 4h^{\mu5}K_{\mu5; \rho5}h^{\rho5}
	+ 2h^{\mu5}K_{\mu5; \phi}\phi \nonumber\\
& 	\hspace{2em}
	+ \phi K_{\phi; \rho\sigma}h^{\rho\sigma}
	+ 2\phi K_{\phi; \rho5}h^{\rho5}
	+ \phi K_{\phi; \phi}\phi\biggr\}, \label{eq:gravity06}
\end{align}
where
\begin{subequations}
\begin{align}
K_{\mu\nu; \rho\sigma}
&= 	- \frac{1}{2}
	(\eta_{\mu\rho}\partial_{\nu}\partial_{\sigma}
	+ \eta_{\mu\sigma}\partial_{\nu}\partial_{\rho}
	+ \eta_{\nu\rho}\partial_{\mu}\partial_{\sigma}
	+ \eta_{\nu\sigma}\partial_{\mu}\partial_{\rho}) \nonumber\\
& 	\hspace{1em}
	+ \eta_{\mu\nu}\partial_{\rho}\partial_{\sigma}
	+ \eta_{\rho\sigma}\partial_{\mu}\partial_{\nu} \nonumber\\
& 	\hspace{1em}
	+ \frac{1}{2}
	(\eta_{\mu\rho}\eta_{\nu\sigma}
	+ \eta_{\mu\sigma}\eta_{\nu\rho}
	- 2\eta_{\mu\nu}\eta_{\rho\sigma}) \nonumber\\
&	\hspace{8em}
	\times
	(\Box + (\partial_{z} + 3A')\partial_{z}), \label{eq:gravity07a}
\end{align}
\begin{align}
K_{\mu\nu; \rho5}
&= 	- \frac{1}{2}
	(\eta_{\mu\rho}\partial_{\nu}
	+ \eta_{\nu\rho}\partial_{\mu}
	- 2\eta_{\mu\nu}\partial_{\rho})
	(\partial_{z} + 3A'), \label{eq:gravity07b}\\
K_{\mu\nu; \phi}
&= 	\frac{3}{2}
	\eta_{\mu\nu}
	(\partial_{z} + 3A')(\partial_{z} + 2A'), \label{eq:gravity07c}\\
K_{\mu5; \rho\sigma}
&= 	- \frac{1}{2}
	(\eta_{\mu\rho}\partial_{\sigma}
	+ \eta_{\mu\sigma}\partial_{\rho}
	- 2\eta_{\rho\sigma}\partial_{\mu})
	\partial_{z}, \label{eq:gravity07d}\\
K_{\mu5; \rho5}
&= 	- \frac{1}{2}(\partial_{\mu}\partial_{\rho} - \eta_{\mu\rho}\Box), \label{eq:gravity07e}\\
K_{\mu5; \phi}
&= 	- \frac{3}{2}\partial_{\mu}(\partial_{z} + 2A'), \label{eq:gravity07f}\\
K_{\phi; \rho\sigma}
&= 	\frac{3}{2}\eta_{\rho\sigma}(\partial_{z} + A')\partial_{z}, \label{eq:gravity07g}\\
K_{\phi; \rho5}
&= 	- \frac{3}{2}\partial_{\rho}(\partial_{z} + A'), \label{eq:gravity07h}\\
K_{\phi; \phi}
&= 	\frac{3}{2}\Box
	- 3(\partial_{z} + A')(\partial_{z} + 2A'). \label{eq:gravity07i}
\end{align}
\end{subequations}
These expressions
\footnote{%
The normalization is different from \cite{LNOSS1:2007} by $1/2$ to canonicalize the operator $K_{\mu\nu;\rho\sigma}$.
}
 are also found in \cite{CR:2005,GPP:2005} and consistent with the Einstein equations \cite{CGGP:2004}.

One can easily verify that the action is invariant under infinitesimal general coordinate transformations
\begin{equation}
x^{M} \mapsto {\hat x}^{M} = x^{M} + \xi^{M}, \label{eq:gravity08}
\end{equation}
which transform the fluctuations at the linearized level as
\begin{subequations}
\begin{align}
h_{\mu\nu}
&\mapsto
{\hat h}_{\mu\nu}
= 	h_{\mu\nu}
	- \partial_{\mu}\xi_{\nu} - \partial_{\nu}\xi_{\mu}
	- \eta_{\mu\nu}(\partial_{z} + 3A')\xi_{5}, \label{eq:gravity09a}\\
h_{\mu5}
&\mapsto
{\hat h}_{\mu5}
= 	h_{\mu5}
	- \partial_{z}\xi_{\mu}
	- \partial_{\mu}\xi_{5}, \label{eq:gravity09b}\\
\phi
&\mapsto
{\hat \phi}
= 	\phi
	- 2(\partial_{z} + A')\xi_{5}. \label{eq:gravity09c}
\end{align}
\end{subequations}

\noindent
{\bf Supersymmetric Structure:}
For the later discussion we here briefly summarize the quantum mechanical supersymmetric structure of the theory.
As shown in ref.\cite{LNOSS1:2007} the fluctuations are expanded as
\begin{subequations}
\begin{align}
h_{\mu\nu}(x,z)
&= 	\sum_{n=0}^{\infty}h^{(n)}_{\mu\nu}(x)f^{(n)}(z), \label{eq:gravity10a}\\
h_{\mu5}(x,z)
&= 	\sum_{n=1}^{\infty}h^{(n)}_{\mu5}(x)g^{(n)}(z), \label{eq:gravity10b}\\
\phi(x,z)
&= 	\sum_{n=0}^{\infty}\phi^{(n)}(x)k^{(n)}(z), \label{eq:gravity10c}
\end{align}
\end{subequations}
where we have excluded the vector zero-mode $g^{(0)}$ from the expansion \eqref{eq:gravity10b}, since as we shall see immediately it will be incompatible with the supersymmetric structure.

\newpage
The orthonormal sets $\{f^{(n)}\}_{n=0}^{\infty}$ $\{g^{(n)}\}_{n=1}^{\infty}$ and $\{k^{(n)}\}_{n=0}^{\infty}$ are defined as the eigenfunctions of the Schr\"odinger-like equations
\begin{widetext}
\begin{subequations}
\begin{align}
\begin{bmatrix}
-(\partial_{z} + 3A')\partial_{z} 	& 0 \\
0 							& -\partial_{z}(\partial_{z} + 3A')
\end{bmatrix}
\begin{bmatrix}
f^{(n)} \\
g^{(n)}
\end{bmatrix}
&=
m_{n}^{2}
\begin{bmatrix}
f^{(n)} \\
g^{(n)}
\end{bmatrix}, \label{eq:gravity11a}\\
\begin{bmatrix}
-(\partial_{z} + 2A')(\partial_{z} + A') 	& 0 \\
0 								& -(\partial_{z} + A')(\partial_{z} + 2A')
\end{bmatrix}
\begin{bmatrix}
g^{(n)} \\
k^{(n)}
\end{bmatrix}
&=
m_{n}^{2}
\begin{bmatrix}
g^{(n)} \\
k^{(n)}
\end{bmatrix}, \label{eq:gravity11b}
\end{align}
\end{subequations}
\end{widetext}
with the boundary conditions
\begin{equation}
\partial_{z}f^{(n)}(z_{i}) = 0, \quad
g^{(n)}(z_{i}) = 0, \quad
(\partial_{z} + 2A')k^{(n)}(z_{i}) = 0, \label{eq:gravity12}
\end{equation}
where $z_{i} = 0$ or $z_{c}$.
Note that the background solution enables us to factorize the Hamiltonian for the vector-mode $g^{(n)}$ in two different ways (except on the boundaries); $-\partial_{z}(\partial_{z} + 3A') = -(\partial_{z} + 2A')(\partial_{z} + A')$.

These eigenfunctions are mutually related through
\begin{subequations}
\begin{align}
\begin{bmatrix}
0 			& -(\partial_{z} + 3A') \\
\partial_{z} 	& 0
\end{bmatrix}
\begin{bmatrix}
f^{(n)} \\
g^{(n)}
\end{bmatrix}
&=
m_{n}
\begin{bmatrix}
f^{(n)} \\
g^{(n)}
\end{bmatrix}, \label{eq:gravity13a}\\
\begin{bmatrix}
0 			& -(\partial_{z} + 2A') \\
\partial_{z}+A'	& 0
\end{bmatrix}
\begin{bmatrix}
g^{(n)} \\
k^{(n)}
\end{bmatrix}
&=
m_{n}
\begin{bmatrix}
g^{(n)} \\
k^{(n)}
\end{bmatrix}. \label{eq:gravity13b}
\end{align}
\end{subequations}
Now it is clear that there is no nontrivial solution of the vector zero-mode $g^{(0)}$ which has to satisfy the two different linear differential equations
\begin{align}
-(\partial_{z} + 3A')g^{(0)} = 0 = (\partial_{z} + A')g^{(0)}.
\end{align}
We have already included this fact into the expansion \eqref{eq:gravity10b}.

It should be noted that the eigenfunctions satisfy the identity
\begin{align}
-2A'(z)g^{(n)}(z) = m_{n}\bigl(f^{(n)}(z) + k^{(n)}(z)\bigr), \quad n\in{\mathbb Z}_{>0}, \label{eq:gravity14}
\end{align}
which plays an essential role in the proof of gauge-independence of one graviton exchange amplitude in the $R_{\xi}$ gauge; see Section \ref{sec:amplitude}.
Now we introduce the following $2\times2$ matrix operators
\begin{subequations}
\begin{align}
H
&= 	\begin{bmatrix}
	-(\partial_{z} + 3A')\partial_{z} 	& 0 \\
	0 							& -\partial_{z}(\partial_{z} + 3A')
	\end{bmatrix}, \nonumber\\
Q_{1}
&= 	\begin{bmatrix}
	0 			& -(\partial_{z} + 3A') \\
	\partial_{z} 	& 0
	\end{bmatrix}, \quad
(-1)^{F}
= 	\begin{bmatrix}
	1 	& 0 \\
	0 	& -1
	\end{bmatrix}, \label{eq:gravity15a}\\
{\bar H}
&= 	\begin{bmatrix}
	-(\partial_{z} + 2A')(\partial_{z}+ A') 	& 0 \\
	0 								& -(\partial_{z} + A')(\partial_{z} + 2A')
	\end{bmatrix}, \nonumber\\
{\bar Q}_{1}
&= 	\begin{bmatrix}
	0 				& -(\partial_{z} + 2A') \\
	\partial_{z} + A' 	& 0
	\end{bmatrix}, \quad
(-1)^{\bar F}
= 	\begin{bmatrix}
	1 	& 0 \\
	0 	& -1
	\end{bmatrix}, \label{eq:gravity15b}
\end{align}
\end{subequations}
which satisfy the {\em two} ${\cal N} = 2$ superalgebras
\begin{subequations}
\begin{align}
& 	\{Q_{a}, Q_{b}\}
= 	2\delta_{ab}H, \quad
	[H, Q_{a}] = [H, (-1)^{F}]
= 	0, \nonumber\\
&	\{(-1)^{F}, Q_{a}\}
= 	0, \label{eq:gravity16a}\\
& 	\{{\bar Q}_{a}, {\bar Q}_{b}\}
= 	2\delta_{ab}{\bar H}, \quad
	[{\bar H}, {\bar Q}_{a}] = [{\bar H}, (-1)^{\bar F}]
= 	0, \nonumber\\
& 	\{(-1)^{\bar F}, {\bar Q}_{a}\}
= 0, \quad
a,b = 1,2, \label{eq:gravity16b}
\end{align}
\end{subequations}
where $Q_{2} = i(-1)^{F}Q_{1}$ and ${\bar Q}_{2} = i(-1)^{\bar F}{\bar Q}_{1}$.

We emphasize that \eqref{eq:gravity12} is the {\em unique} boundary conditions to ensure the Hamiltonians $H$ and ${\bar H}$ to be isospectral (except for the zero modes), or the supersymmetries in the mass spectrum.
Indeed the hermiticity of the Hamiltonians and supercharges is ensured if and only if the boundary conditions \eqref{eq:gravity12} are imposed (see ref.\cite{LNOSS1:2007}).

\vskip 2ex
\noindent
{\bf Unitary Gauge}:
Next we will identify the particle content of the theory.
To this end we first note that using the mode decompositions \eqref{eq:gravity10a} -- \eqref{eq:gravity10c} the linearized general coordinate transformations \eqref{eq:gravity09a} -- \eqref{eq:gravity09c} read
\begin{subequations}
\begin{align}
\delta h^{(n)}_{\mu\nu}
&= 	- \partial_{\mu}\xi^{(n)}_{\nu}
	- \partial_{\nu}\xi^{(n)}_{\mu}
	+ m_{n}\eta_{\mu\nu}\xi^{(n)}_{5},
	\hspace{1em} n\in\mathbb{Z}_{\geq0}, \label{eq:gravity17a}\\
\delta h^{(n)}_{\mu5}
&= 	- m_{n}\xi^{(n)}_{\mu} - \partial_{\mu}\xi^{(n)}_{5},
	\hspace{1em} n\in\mathbb{Z}_{>0}, \label{eq:gravity17b}\\
\delta\phi^{(n)}
&= 	- 2m_{n}\xi^{(n)}_{5}, \hspace{1em} n\in\mathbb{Z}_{\geq0}, \label{eq:gravity17c}
\end{align}
\end{subequations}
where the gauge freedom $\xi_{M}$ are assumed to be expanded as
\begin{equation}
\xi_{\mu}(x,z)
= 	\sum_{n=0}^{\infty}\xi_{\mu}^{(n)}(x)f^{(n)}(z), \quad
\xi_{5}(x,z)
= 	\sum_{n=1}^{\infty}
	\xi_{5}^{(n)}(x)g^{(n)}(z). \label{eq:gravity18}
\end{equation}
If we move on to the coordinate frame by choosing
\begin{align}
\xi_{5}^{(n)}
= 	\frac{1}{2m_{n}}\phi^{(n)}, \quad
\xi_{\mu}^{(n)}
= 	\frac{1}{m_{n}}h_{\mu5}^{(n)}
	- \frac{1}{2m_{n}^{2}}\partial_{\mu}\phi^{(n)}, \quad
n\in{\mathbb Z}_{>0}, \label{eq:gravity19}
\end{align}
$\phi^{(n)}$ and $h_{\mu5}^{(n)}$
 with $n>0$ are all gauged away, leaving only
\begin{subequations}
\begin{align}
{\hat h}_{\mu\nu}^{(n)}
&= 	h_{\mu\nu}^{(n)}
	- \frac{1}{m_{n}}\bigl(\partial_{\mu}h_{\nu}^{(n)} + \partial_{\nu}h_{\mu}^{(n)}\bigr) \nonumber\\
&	\hspace{1em}
	+ \frac{1}{2}\left(\eta_{\mu\nu} + 2\frac{\partial_{\mu}\partial_{\nu}}{m_{n}^{2}}\right)\phi^{(n)}, \quad
	n\in\mathbb{Z}_{>0}, \label{eq:gravity20a}\\
{\hat h}_{\mu\nu}^{(0)}
&= 	h_{\mu\nu}^{(0)}, \quad
{\hat \phi}^{(0)}
= 	\phi^{(0)}, \label{eq:gravity20b}
\end{align}
\end{subequations}
so that the quadratic action becomes
\footnote{%
The eigenfunctions are normalized as
\begin{align}
M^{3}\!\!\int_{0}^{z_{c}}\!\!\!{\rm d}z~{\rm e}^{3A}f^{(n)}(z)f^{(m)}(z)
& = M_{\rm Pl}^{2}\delta_{nm}, \nonumber\\
M^{3}\!\!\int_{0}^{z_{c}}\!\!\!{\rm d}z~{\rm e}^{3A}g^{(n)}(z)g^{(m)}(z)
&= M_{\rm Pl}^{2}\delta_{nm}, \nonumber\\
M^{3}\!\!\int_{0}^{z_{c}}\!\!\!{\rm d}z~{\rm e}^{3A}k^{(n)}(z)k^{(m)}(z)
&= M_{\rm Pl}^{2}\delta_{nm}. \nonumber\\
\end{align}
}
\begin{align}
S^{(2)}
= 	M_{\rm Pl}^{2}\!\!\int{\rm d}^{4}x
	\left\{
	\sum_{n=0}^{\infty}\frac{1}{4}
	{\hat h}^{(n)\mu\nu}K^{(n)}_{\mu\nu; \rho\sigma}{\hat h}^{(n)\rho\sigma}
	+ \frac{3}{8}{\hat\phi}^{(0)}\Box{\hat\phi}^{(0)}\right\}, \label{eq:gravity20}
\end{align}
where $M_{\rm Pl}$ is the (5d averaged) 4d Planck mass scale defined as
\begin{align}
M_{\rm Pl}^{2}
= 	M^{3}\int_{0}^{z_{c}}\!\!\!{\rm d}z~{\rm e}^{3A}
= 	\frac{M^{3}}{2k}(1 - {\rm e}^{-2\pi kR}), \label{eq:gravity21}
\end{align}
which reduces to $M_{\rm Pl}^{2} \xrightarrow{\Lambda \to 0} \pi RM^{3}$ as it should, and
\begin{align}
K^{(n)}_{\mu\nu; \rho\sigma}
&= 	- \frac{1}{2}
	(\eta_{\mu\rho}\partial_{\nu}\partial_{\sigma}
	+ \eta_{\mu\sigma}\partial_{\nu}\partial_{\rho}
	+ \eta_{\nu\rho}\partial_{\mu}\partial_{\sigma}
	+ \eta_{\nu\sigma}\partial_{\mu}\partial_{\rho}) \nonumber\\
& 	\hspace{1em}
	+ \eta_{\mu\nu}\partial_{\rho}\partial_{\sigma}
	+ \eta_{\rho\sigma}\partial_{\mu}\partial_{\nu} \nonumber\\
& 	\hspace{1em}
	+ \frac{1}{2}
	(\eta_{\mu\rho}\eta_{\nu\sigma}
	+ \eta_{\mu\sigma}\eta_{\nu\rho}
	- 2\eta_{\mu\nu}\eta_{\rho\sigma})
	(\Box - m_{n}^{2}). \label{eq:gravity22}
\end{align}
Notice that the graviton mass appears as the so-called Fierz-Pauli \cite{FP:1939} form, which is the only form that does not introduce ghosts \cite{Nieuwenhuizen:1973fi}.
The four-dimensional field contents are now obvious: the massless graviton $h^{(0)}_{\mu\nu}$, the infinite tower of massive gravitons $h^{(n)}_{\mu\nu}$ with increasing masses $\{m_{n}\}_{n=1}^{\infty}$ and the massless scalar boson $\phi^{(0)}$ known as the radion.
It is worth mentioning at this point that $M_{\rm Pl}$ is {\em not} directly related to the Newton constant $G_{N}$; $M_{\rm pl}$ is just introduced as the coefficient of the massless graviton Lagrangian, whereas $G_{N}$ has to be determined as the coefficient of the inverse square law part of the static gravitational force, which now includes the massless radion's contribution as well as the massless graviton's one.

%%%%% SUB-SECTION 3.1 %%%%%
\subsection{Gauge-Fixed Action to Quadratic Order} \label{subsec:gravity_gauge-fix}
Once fixed the action to the unitary gauge, all the fictitious massive vector- and scalar-modes would disappear from the dimensional reduced action so that we could not see the cancelation mechanism of the would-be NG bosons which appear in the other gauges, nor the expected SUSY-like symmetries between the unphysical degrees of freedom with the same gauge-dependent mass-squared.
In view of this we have to establish the one-parameter family of gauge choices analogous to the $R_{\xi}$ gauge in the previous section.
However, it is not straightforward to find such a gauge due to the appearance of three undesirable quadratic mixing terms.

Our strategy to find out the appropriate gauge-fixing function similar to the $R_{\xi}$ gauge in spontaneously broken gauge theories is based on the observations:

(1) In 4d general relativity the so-called harmonic gauge, which is the counterpart of the Lorentz gauge in ordinary  unbroken gauge theories, is just the linearized version of $\partial_{\mu}(\sqrt{-g}g^{\mu\nu}) = 0$.
Therefore we could expect that the harmonic gauge conditions for the warped Randall-Sundrum metric would also be given by something like the conditions $\partial_{M}(\sqrt{-G}G^{MN}) = 0$ to linear order of ${\bar h}_{MN}$.

(2) In order to fix the gauge arbitrariness $\xi_{M}$ we have to impose five independent conditions.
The gauge-fixing function for $\xi_{\mu}$ must be expanded by the orthonormal set with the same boundary condition to $\xi_{\mu}$, i.e. $\{f^{(n)}(z)\}_{n=0}^{\infty}$, whereas that for $\xi_{5}$ must  be expanded by $\{g^{(n)}(z)\}_{n=1}^{\infty}$.
As we shall see immediately, this is a key observation to select the terms adaptable to the gauge-fixing function from the linearized $\partial_{M}(\sqrt{-G}G^{MN})$.

In what follows we first investigate the harmonic gauge (or 't Hooft-Feynman gauge from the four-dimensional spontaneously broken point of view) for the warped gravity and then establish the one-parameter family of gauge choices analogous to the $R_{\xi}$ gauge in spontaneously broken gauge theories.
The analysis presented below is based on the knowledge of gauge-fixing in 4d general relativity, which is briefly summarized in Appendix \ref{appendix:vDVZ} together with a brief review of the vDVZ-discontinuity.

\vskip 2ex
\noindent
{\bf 't Hooft-Feynman Gauge}:
As in 4d general relativity, we will first calculate $\partial_{M}\left(\sqrt{- G}G^{MN}\right)$ to linear order:
\begin{subequations}
\begin{align}
\partial_{M}\left(\sqrt{- G}G^{M\nu}\right)
&= 	{\rm e}^{3A}
	\left[- \partial_{\mu}h^{\mu\nu}
	+ \frac{1}{2}\partial^{\nu}h
	- \left(\partial_{z} + 3A'\right)h^{\nu5}\right] \nonumber\\
&	\hspace{1em}
	+ O({\bar h}^{2}), \label{eq:gravity23a}\\
\partial_{M}\left(\sqrt{- G}G^{M5}\right)
&= 	{\rm e}^{3A}
	\biggl[3A'
	+ \frac{3}{2}A'\left(h - \phi\right)
	+ \frac{1}{2}\partial_{z}h
	- \partial_{\mu}h^{\mu 5} \nonumber\\
&	\hspace{1em}
	- \frac{3}{2}\left(\partial_{z} + 2A'\right)\phi\biggr]
	+ O({\bar h}^{2}), \label{eq:gravity23b}
\end{align}
\end{subequations}
where $h = \eta^{\mu\nu}h_{\mu\nu}$.
First look at the right hand side (RHS) of Eq.\eqref{eq:gravity23a}.
The three terms in the brace can be expanded in only $f^{(n)}$-modes so that it seems to be adopted as a gauge-fixing term.
Next look at RHS of Eq.\eqref{eq:gravity23b}. 
Since we are now interested in gauge-fixing for quadratic part of the action, the first term $3A'$ should be ignored, because it gives 0th and 1st order of the fields.
The last three terms in the brace can be expanded in $g^{(n)}$-modes, however, the second term can not.
Thus, for the present, we will adopt only the last three terms for gauge-fixing and examine whether it works well or not.

Our modified harmonic gauge conditions are therefore
\begin{subequations}
\begin{align}
F_{\mu}[h]
&= 	- \partial^{\lambda}h_{\lambda\mu}
	+ \frac{1}{2}\partial_{\mu}h
	- \left(\partial_{z} + 3A'\right)h_{\mu5}
= 	0, \label{eq:gravity24a}\\
F_{5}[h]
&= 	\frac{1}{2}\partial_{z}h
	- \partial^{\mu}h_{\mu 5}
	- \frac{3}{2}\left(\partial_{z} + 2A'\right)\phi
= 	0. \label{eq:gravity24b}
\end{align}
\end{subequations}
Let us first examine how to implement these gauge-fixing conditions.
To see this, it is sufficient to calculate $\delta F_{\mu}[h]$ and $\delta F_{5}[h]$.
Using the transformation law of the fluctuations at KK-mode level we get
\begin{subequations}
\begin{align}
\delta F_{\mu}[h]
&= 	\sum_{n=0}^{\infty}
	\left[
	\left(\Box - m_{n}^{2}\right)\xi_{\mu}^{(n)}
	\right]f^{(n)}, \label{eq:gravity25a}\\
\delta F_{5}[h]
&= 	\sum_{n=1}^{\infty}
	\left[
	\left(\Box - m_{n}^{2}\right)\xi_{5}^{(n)}
	\right]g^{(n)}. \label{eq:gravity25b}
\end{align}
\end{subequations}
It is obvious that one can always impose the conditions \eqref{eq:gravity24a} and \eqref{eq:gravity24b} by choosing $\xi^{M}$ as one of solutions to the following differential equations for each $n$:
\begin{subequations}
\begin{align}
(\Box - m_{n}^{2})\xi_{\nu}^{(n)}
&= 	\partial^{\mu}h_{\mu\nu}^{(n)}
	- \frac{1}{2}\partial_{\nu}h^{(n)}
	- m_{n}h_{\nu5}^{(n)}, 
	\quad n\in\mathbb{Z}_{\geq0}, \label{eq:gravity26a}\\
(\Box - m_{n}^{2})\xi_{5}^{(n)}
&= 	- \frac{1}{2}m_{n}h^{(n)}
	+ \partial^{\mu}h_{\mu5}^{(n)}
	- \frac{3}{2}m_{n}\phi^{(n)},
	\quad n\in\mathbb{Z}_{>0}. \label{eq:gravity26b}
\end{align}
\end{subequations}
These are the analogue of Eq.\eqref{eq:massless08}.
Notice that when one impose the gauge-fixing conditions \eqref{eq:gravity24a} and \eqref{eq:gravity24b}, however, there still remains gauge freedom generated by on-shell parameters $\xi_{\mu}^{(n)}$ and $\xi_{5}^{(n)}$ which satisfy $(\Box - m_{n}^{2})\xi_{\mu}^{(n)} = 0$ and $(\Box - m_{n}^{2})\xi_{5}^{(n)} = 0$.
This is also the analogue of residual on-shell gauge freedom in 4d general relativity.

Adding to the action \eqref{eq:gravity06} a gauge-fixing term
\begin{equation}
{\cal L}_{\rm GF}^{\xi = 1}
=	M^{3}{\rm e}^{3A}
	\left\{
	- \frac{1}{2}\bigl(F_{\mu}[h]\bigr)^{2}
	- \frac{1}{2}\bigl(F_{5}[h]\bigr)^{2}\right\}, \label{eq:gravity27}
\end{equation}
\begin{widetext}
we get
\begin{align}
S_{\xi = 1}^{(2)}
&= 	M^{3}\!\!\int{\rm d}^{4}x\int_{0}^{z_{c}}\!\!\!{\rm d}z~{\rm e}^{3A}
	\biggl\{
	\frac{1}{4}h^{\mu\nu}
	\left[\frac{1}{2}(\eta_{\mu\rho}\eta_{\nu\sigma}
	+ \eta_{\mu\sigma}\eta_{\nu\rho}
	- \eta_{\mu\nu}\eta_{\rho\sigma})
	\bigl(\Box + (\partial_{z} + 3A')\partial_{z}\bigr)\right]
	h^{\rho\sigma}. \nonumber\\
&	\hspace{9em}
	+ \frac{1}{2}h^{\mu5}\left[
	\eta_{\mu\nu}\bigl(\Box + \partial_{z}(\partial_{z} + 3A')\bigr)\right]
	h^{\nu5}
	+ \frac{3}{8}\phi\left[
	\Box + (\partial_{z} + A')
	(\partial_{z} + 2A')\right]
	\phi\biggr\}. \label{eq:gravity28}
\end{align}
The dimensional reduced action is
\begin{align}
S^{(2)}_{\xi=1}
&= 	M_{\rm Pl}^{2}\int{\rm d}^{4}x
	\biggl\{\sum_{n=0}^{\infty}
	\frac{1}{4}h^{\mu\nu(n)}
	\left[\frac{1}{2}(\eta_{\mu\rho}\eta_{\nu\sigma}
		+ \eta_{\mu\sigma}\eta_{\nu\rho}
		- \eta_{\mu\nu}\eta_{\rho\sigma})
		(\Box - m_{n}^{2})\right]
	h^{\rho\sigma(n)} \nonumber\\
&	\hspace{5em}
	+ \sum_{n=1}^{\infty}
	\frac{1}{2}h^{\mu5(n)}
	[\eta_{\mu\nu}(\Box - m_{n}^{2})]
	h^{\nu5(n)}
	+ \sum_{n=0}^{\infty}
	\frac{3}{8}\phi^{(n)}
	[\Box - m_{n}^{2}]
	\phi^{(n)} \biggr\}. \label{eq:gravity29}
\end{align}
It should be emphasized the first line suggests that the residue of the graviton propagator is common to {\em any} $n\in{\rm Z}_{\geq0}$ so that one may expect in the 't Hooft-Feynman gauge it may be possible to evade the vDVZ discontinuity, which arises in the unitary gauge.
For further discussion, see Section \ref{sec:amplitude}.
\end{widetext}

\noindent
{\bf $\bm R_{\xi}$ Gauge}:
So far we have used the notation ${\cal L}^{\xi=1}_{\rm GF}$ and $S^{(2)}_{\xi=1}$,
which implies that, as in gauge theories, our 't Hooft-Feynman gauge just corresponds to the case $\xi = 1$ in somewhat underlying $R_{\xi}$ gauge, however, no mention has been made of how we have inserted the gauge parameter $\xi$ into the action.
Actually, it is highly non-trivial to insert $\xi$ into the gauge-fixing Lagrangian \eqref{eq:gravity27}.
To see the difficulty, consider the arithmetic
\begin{align}
& 	\frac{1}{2\xi}(x + ay + bz)^{2} \nonumber\\
&= 	\frac{1}{2\xi}(x^{2} + b^{2}y^{2} + c^{2}z^{2}
	+ 2axy + 2bxz + 2abyz). \label{eq:gravity30}
\end{align}
Since we already know the 't Hooft-Feynman gauge, which corresponds to $\xi = a = b = 1$, we now want to choose the coefficients $a$ and $b$ to make the three cross terms independent of $\xi$: $a/\xi = b/\xi = (ab)/\xi = 1$.
But obviously it is not possible.
At first glance, it seems to be impossible to eliminate the three undesirable mixing terms by a single gauge parameter $\xi$.
However, we can construct the one-parameter family of gauge choices analogous to the $R_{\xi}$ gauge in spontaneously broken gauge theories with the help of the boundary conditions \eqref{eq:gravity12}.

One of the crucial keys is that we have two squared brackets in \eqref{eq:gravity27}, $\bigl(F_{\mu}[h]\bigr)^{2}$ and $\bigl(F_{5}[h]\bigr)^{2}$, the former fixes the four-dimensional gauge arbitrariness $\xi_{\mu}$ and the latter the extra-dimensional one $\xi_{5}$.
Since the scalar field $\phi$ only appears in the latter bracket, we have to insert $\xi$ as a coefficient of $\phi$ to cancel out by the overall factor $1/(2\xi)$.
Then the cross term between $h$ and $h^{\mu5}$ in the latter bracket come to depend on $\xi$.
However, we can make all the cross terms independent of $\xi$ by inserting it into the former bracket as following combination:
\begin{align}
{\cal L}_{\rm GF}
=	M^{3}{\rm e}^{3A}
	\left\{
	- \frac{1}{2}\bigl(F^{R_{\xi}}_{\mu}[h]\bigr)^{2}
	- \frac{1}{2}\bigl(F^{R_{\xi}}_{5}[h]\bigr)^{2}\right\}, \label{eq:gravity31}
\end{align}
with
\begin{subequations}
\begin{align}
F^{R_{\xi}}_{\mu}[h]
&= 	-\partial^{\lambda}h_{\lambda\mu}
	+ \frac{1}{2}\left(2 - \frac{1}{\xi}\right)\partial_{\mu}h
	- \xi(\partial_{z} + 3A')h_{\mu5}, \label{eq:gravity32a}\\
F^{R_{\xi}}_{5}[h]
&= 	\frac{1}{2}\partial_{z}h
	- \partial^{\mu}h_{\mu5}
	- \frac{3\xi}{2}(\partial_{z} + 2A')\phi. \label{eq:gravity32b}
\end{align}
\end{subequations}
Upon integrating by parts, one can easily check that the cross term between $h$ and $h^{\mu5}$ is just the same as \eqref{eq:gravity27} with the help of the boundary conditions \eqref{eq:gravity12}.
Thus \eqref{eq:gravity31} correctly eliminates the undesirable mixing terms.

\begin{widetext}
\noindent
{\bf Gauge-Fixed Action to Quadratic Order}:
Adding to the action \eqref{eq:gravity06} a gauge-fixing term \eqref{eq:gravity31} we get
\begin{align}
S_{R_{\xi}}^{(2)}
&= 	M^{3}\!\!\int{\rm d}^{4}x\int_{0}^{z_{c}}\!\!\!{\rm d}z~{\rm e}^{3A}
	\biggl\{\frac{1}{4}h^{\mu\nu}K_{\mu\nu; \rho\sigma}^{R_{\xi}}h^{\rho\sigma}
	+ \frac{1}{2}h^{\mu5}K_{\mu; \rho}^{R_{\xi}}h^{\rho5}
	+ \frac{3}{8}\phi K_{\phi; \phi}^{R_{\xi}}\phi\biggr\}, \label{eq:gravity33}
\end{align}
where
\begin{subequations}
\begin{align}
K_{\mu\nu; \rho\sigma}^{R_{\xi}}
&= 	- \frac{1}{2}\left(1 - \frac{1}{\xi}\right)
	(\eta_{\mu\rho}\partial_{\nu}\partial_{\sigma}
	+ \eta_{\mu\sigma}\partial_{\nu}\partial_{\rho}
	+ \eta_{\nu\rho}\partial_{\mu}\partial_{\sigma}
	+ \eta_{\nu\sigma}\partial_{\mu}\partial_{\rho}) \nonumber\\
& 	\hspace{1em}
	+ \left(1 - \frac{1}{\xi}\right)^{2}
	\eta_{\mu\nu}\partial_{\rho}\partial_{\sigma}
	+ \left(1 - \frac{1}{\xi}\right)^{2}
	\eta_{\rho\sigma}\partial_{\mu}\partial_{\nu} \nonumber\\
& 	\hspace{1em}
	+ \frac{1}{2}
	\Biggl(\eta_{\mu\rho}\eta_{\nu\sigma}
	+ \eta_{\mu\sigma}\eta_{\nu\rho}
	- 2\left(1 - \frac{1}{2\xi}\left(2 - \frac{1}{\xi}\right)^{2}\right)
	\eta_{\mu\nu}\eta_{\rho\sigma}\Biggr)\Box \nonumber\\
& 	\hspace{1em}
	+ \frac{1}{2}
	\left(\eta_{\mu\rho}\eta_{\nu\sigma}
	+ \eta_{\mu\sigma}\eta_{\nu\rho}
	- \left(2 - \frac{1}{\xi}\right)\eta_{\mu\nu}\eta_{\rho\sigma}\right)
	(\partial_{z} + 3A')\partial_{z}, \label{eq:gravity34a}\\
K_{\mu; \rho}^{R_{\xi}}
&=	\eta_{\mu\rho}[\Box + \xi\partial_{z}(\partial_{z} + 3A')]
	- \left(1 - \frac{1}{\xi}\right)
	\partial_{\mu}\partial_{\rho}, \label{eq:gravity34b}\\
K_{\phi; \phi}^{R_{\xi}}
&= 	\Box + (3\xi - 2)(\partial_{z} + A')(\partial_{z} + 2A'). \label{eq:gravity34c}
\end{align}
\end{subequations}
\end{widetext}
Now it is straightforward to compute the propagators in the $R_{\xi}$ gauge.
The results are summarized in Appendix \ref{appendix:propagator}.
As in a massive gauge boson propagator of spontaneously broken gauge theories, the massive graviton propagator can be written into the sum of  the gauge-independent and -dependent part separately.
The gauge-independent part corresponds to the transverse-traceless part of $h_{\mu\nu}$, as it should.
The gauge-dependent part can further be decomposed into the sum of the three independent parts;
the transverse gauge freedom part $\Delta^{({\rm T})}$, the longitudinal gauge freedom part $\Delta^{({\rm L})}$ and the trace part $\Delta^{({\rm tr})}$.
$\Delta^{({\rm T})}$ has the same simple pole position to the transverse part of the vector propagator.
$\Delta^{({\rm L})}$ and $\Delta^{({\rm tr})}$ have the same simple pole positions to the longitudinal part of the vector propagator and the scalar propagator, respectively.
In view of this fact we expect that as in the case of the pure Abelian gauge theory, these spurious degrees of freedom have its partners with the same mass-squared, and consist of multiplets under SUSY-like transformations generated by the supercharges.

To this end we decompose the graviton and vector fields into the following polarization states
\begin{subequations}
\begin{align}
h_{\mu\nu}
&= 	h^{({\rm TT})}_{\mu\nu}
	+ \partial_{\mu}h^{({\rm T})}_{\nu} + \partial_{\nu}h^{({\rm T})}_{\mu}
	+ 2\partial_{\mu}\partial_{\nu}h^{({\rm L})}
	+ \eta_{\mu\nu}h^{({\rm tr})}, \label{eq:gravity35a}\\
h_{\mu5}
&= 	h^{({\rm T})}_{\mu5}
	+ \partial_{\mu}h^{({\rm L})}_{5}, \label{eq:gravity35b}
\end{align}
\end{subequations}
where $h^{({\rm TT})}_{\mu\nu}$ is the transverse-traceless mode (5 DOF), $h^{({\rm T})}_{\mu}$ the transverse gauge freedom (3 DOF), $h^{({\rm L})}$ the longitudinal gauge freedom (1 DOF), $h^{({\rm tr})}$ the ``trace'' piece (1 DOF), $h^{({\rm T})}_{\mu}$ the transverse mode (3 DOF) and $h^{({\rm L})}_{5}$ the longitudinal mode (1 DOF).
Notice that these decompositions are justified only for the case that $h_{\mu\nu}$ and $h_{\mu5}$ are both massive.
Indeed the massless tensor decomposition are more complicated; see for example Appendix of \cite{CLP:2005}.
However, since we are now interested in the gauge-dependent mass terms of the would-be NG bosons, we do not need the massless tensor and vector decompositions.

%%%%% SUB-SECTION 3.2 %%%%%
\begin{widetext}
\subsection{Residual Symmetries} \label{subsec:gravity_sym}
Inserting the decompositions \eqref{eq:gravity35a}, \eqref{eq:gravity35b} into \eqref{eq:gravity33} we get
\begin{align}
S^{(2)}_{R_{\xi}}
&= 	M^{3}\!\!\int{\rm d}^{4}x\int_{0}^{z_{c}}\!\!\!{\rm d}z~{\rm e}^{3A}
	\biggl\{
	\frac{1}{2}{\hat h}^{({\rm TT})\mu\nu}
	\bigl[\Box + (\partial_{z} + 3A')\partial_{z}\bigr]
	{\hat h}^{({\rm TT})}_{\mu\nu} \nonumber\\
& 	\hspace{1em}
	+ \frac{1}{2}
		\raisebox{2.5ex}{$t$}\!
		\begin{bmatrix}
		{\hat h}^{({\rm T})\mu} \\
		{\hat h}^{({\rm T})\mu5}
		\end{bmatrix}
		\begin{bmatrix}
		-1 	& 0 \\
		0 	& 1
		\end{bmatrix}
		\left(\Box - \xi
			\begin{bmatrix}
			-(\partial_{z} + 3A')\partial_{z} 	& 0 \\
			0 							& -\partial_{z}(\partial_{z} + 3A')
			\end{bmatrix}\right)
		\begin{bmatrix}
		{\hat h}^{({\rm T})}_{\mu} \\
		{\hat h}^{({\rm T})}_{\mu5}
		\end{bmatrix} \nonumber\\
&	\hspace{1em}
	+ \frac{1}{2}
		\raisebox{2.5ex}{$t$}\!
		\begin{bmatrix}
		{\hat h}^{({\rm L})} \\
		{\hat h}^{({\rm L})}_{5}
		\end{bmatrix}
		\begin{bmatrix}
		1 	& 0 \\
		0 	& -1
		\end{bmatrix}
		\left(\Box - \xi^{2}
			\begin{bmatrix}
			-(\partial_{z} + 3A')\partial_{z} 	& 0 \\
			0 							& -\partial_{z}(\partial_{z} + 3A')
			\end{bmatrix}\right)
		\begin{bmatrix}
		{\hat h}^{({\rm L})} \\
		{\hat h}^{({\rm L})}_{5}
		\end{bmatrix} \nonumber\\
& 	\hspace{1em}
	+ \frac{1}{2}
		\raisebox{2.5ex}{$t$}\!
		\begin{bmatrix}
		{\hat h}^{({\rm tr})} \\
		{\hat \phi}
		\end{bmatrix}
		\begin{bmatrix}
		-1 	& 0\\
		0 	& 1
		\end{bmatrix}
		\left(\Box - (3\xi - 2)
			\begin{bmatrix}
			-(\partial_{z} + 3A')\partial_{z} 	& 0 \\
			0 							& -(\partial_{z} + A')(\partial_{z} + 2A')
			\end{bmatrix}\right)
		\begin{bmatrix}
		{\hat h}^{({\rm tr})} \\
		{\hat \phi}
		\end{bmatrix}\biggr\}, \label{eq:gravity36}
\end{align}
\end{widetext}
where
\footnote
{%
Once again, these definitions \eqref{eq:gravity37} are well-defined only in the region inside (outside) the light-cone and $\xi>0$ ($\xi<0$.)
}
\begin{align}
& 	{\hat h}^{({\rm TT})}_{\mu\nu}
	:= \frac{1}{\sqrt{2}}h^{({\rm TT})}_{\mu\nu}, \quad
	{\hat h}^{({\rm T})}_{\mu}
	:= \sqrt{\frac{\Box}{\xi}}h^{({\rm T})}_{\mu}, \quad
	{\hat h}^{({\rm T})}_{\mu5}
	:=  h^{({\rm T})}_{\mu5}, \nonumber\\
& 	{\hat h}^{({\rm L})}
	:= \frac{1}{\sqrt{\xi^{3}}}\bigl(\Box h^{({\rm L})} - (3\xi - 2)h^{({\rm tr})}\bigr), \quad
	{\hat h}^{({\rm L})}_{5}
	:= \sqrt{\frac{\Box}{\xi}}h^{({\rm L})}_{5}, \nonumber\\
& 	{\hat h}^{({\rm tr})}
	:= \sqrt{3}h^{({\rm tr})}, \quad
	{\hat \phi}
	:= \frac{\sqrt{3}}{2}\phi. \label{eq:gravity37}
\end{align}
This is our final expression for the quadratic action.
We have separated the action to the gauge-independent and -dependent part, just as done in the pure Abelian gauge theory.
It is now obvious that \eqref{eq:gravity36} is invariant under the following three independent global transformations
\begin{subequations}
\begin{align}
\delta%_{\theta_{1}}
	\begin{bmatrix}
	{\hat h}^{({\rm T})}_{\mu} \\
	{\hat h}^{({\rm T})}_{\mu5}
	\end{bmatrix}
&= 	\theta_{1}
	\begin{bmatrix}
	0 			& -(\partial_{z} + 3A') \\
	\partial_{z} 	& 0
	\end{bmatrix}
	\begin{bmatrix}
	{\hat h}^{({\rm T})}_{\mu} \\
	{\hat h}^{({\rm T})}_{\mu5}
	\end{bmatrix}, \label{eq:gravity38a}\\
\delta%_{\theta_{2}}
	\begin{bmatrix}
	{\hat h}^{({\rm L})} \\
	{\hat h}^{({\rm L})}_{5}
	\end{bmatrix}
&= 	\theta_{2}
	\begin{bmatrix}
	0			& -(\partial_{z} + 3A') \\
	\partial_{z} 	& 0
	\end{bmatrix}
	\begin{bmatrix}
	{\hat h}^{({\rm L})} \\
	{\hat h}^{({\rm L})}_{5}
	\end{bmatrix}, \label{eq:gravity38b}\\
\delta%_{\theta_{3}}
	\begin{bmatrix}
	{\hat h}^{({\rm tr})} \\
	{\hat \phi}
	\end{bmatrix}
&= 	\theta_{3}
	\begin{bmatrix}
	0 						& -(\partial_{z} + 3A')(\partial_{z} + 2A') \\
	-(\partial_{z} + A')\partial_{z} 	& 0
	\end{bmatrix}
	\begin{bmatrix}
	{\hat h}^{({\rm tr})} \\
	{\hat \phi}
	\end{bmatrix}. \label{eq:gravity38c}
\end{align}
\end{subequations}
As expected, the transformations \eqref{eq:gravity38a} and \eqref{eq:gravity38b} are generated by supercharge $Q_{1}$ in \eqref{eq:gravity15a}.
The transformation \eqref{eq:gravity38c} is not, however, expressed by a first-order differential operator.
This is because \eqref{eq:gravity38c} is the symmetry transformation between (the unphysical component of) $h_{\mu\nu}$ and $\phi$, which should be generated by product of two supercharges.

Several comments are in order at this stage:

(1) It can be regarded that \eqref{eq:gravity38c} is the so-called higher-derivative supersymmetric transformation.
Indeed, if we introduce the following $2\times 2$ matrix operators
\begin{align}
{\hat H}
&= 	\begin{bmatrix}
	-(\partial_{z} + 3A')\partial_{z} 	& 0 \\
	0 							& -(\partial_{z} + A')(\partial_{z} + 2A')
	\end{bmatrix}, \nonumber\\
{\hat Q}_{1}
&= 	\begin{bmatrix}
	0 						& -(\partial_{z} + 3A')(\partial_{z} + 2A') \\
	-(\partial_{z} + A')\partial_{z} 	& 0
	\end{bmatrix}, \nonumber\\
(-1)^{\hat F}
&= 	\begin{bmatrix}
	1 	& 0 \\
	0 	& -1
	\end{bmatrix}, \label{eq:gravity39}
\end{align}
it is easy to check that these operators satisfy the so-called higher-derivative supersymmetry algebra (HSUSY algebra):
\begin{align}
& 	\{{\hat Q}_{a}, {\hat Q}_{b}\} = 2\delta_{ab}{\hat H}^{2}, \quad
	[{\hat H}, {\hat Q}_{a}] = [{\hat H}, (-1)^{\hat F}] = 0, \nonumber\\
& 	\{(-1)^{\hat F}, {\hat Q}_{a}\} = 0, \quad
	a,b = 1,2, \label{eq:gravity40}
\end{align}
where ${\hat Q}_{2} = i(-1)^{\hat F}{\hat Q}_{1}$.
This is an extended nonlinear superalgebra discussed in \cite{ACDI:1995,AIN:1995,Fernandez:1997,AIN:1999,AST:2001,CIN:2002,AS:2003}.
As is different from the standard superalgebra \eqref{eq:gravity16a}, \eqref{eq:gravity16b}, the supercharges are now the second order differential operators, as they should from the mention above.
The point is once again the refactorization of the Hamiltonian for the vector-mode, $-\partial_{z}(\partial_{z} + 3A') = -(\partial_{z} + 2A')(\partial_{z} + A')$, with the help of the background solution $A'' = (A')^{2}$ (except on the boundaries).

(2) 
There is no residual symmetry generated by ${\bar Q}_{1}$.
One may wonder about the absence of the symmetry generated by ${\bar Q}_{1}$.
However, it is reasonable from the unitarity point of view.
In ordinary spontaneously broken gauge theories, the $\xi$-dependent part of the gauge boson propagator must have as its partners the would-be NG bosons and the Faddeev-Popov ghosts with the same $\xi$-dependent pole so as to cancel out the $\xi$-dependence of any physical amplitudes, and then maintain the unitarity of the theory.
In this case, since we do not need the Faddeev-Popov ghosts, the unphysical polarization states of the massive graviton field (5 DOF) must have as its partners the would-be NG vector (4 DOF) and scalar (1 DOF) bosons.
Thus, we do not have any degrees of freedom to relate by the symmetry transformation generated by ${\bar Q}_{1}$.

(3) 
In special cases $\xi = 1,2$, the four polarization states in \eqref{eq:gravity38b} and \eqref{eq:gravity38c} are happened to be in degenerate with the same gauge-dependent mass-squared.
Hence the action comes to exhibit the following four additional global symmetries:
\begin{subequations}
\begin{align}
\delta%_{\theta_{4}}
	\begin{bmatrix}
	{\hat h}^{({L})} \\
	{\hat h}^{({\rm tr})}
	\end{bmatrix}
&= 	\theta_{4}
	\begin{bmatrix}
	0 	& 1 \\
	1 	& 0
	\end{bmatrix}
	\begin{bmatrix}
	{\hat h}^{({L})} \\
	{\hat h}^{({\rm tr})}
	\end{bmatrix}, \label{eq:gravity41a}\\
\delta%_{\theta_{5}}
	\begin{bmatrix}
	{\hat h}^{({L})} \\
	{\hat \phi}
	\end{bmatrix}
&= 	\theta_{5}
	\begin{bmatrix}
	0 						& (\partial_{z} + 3A')(\partial_{z} + 2A') \\
	-(\partial_{z} + A')\partial_{z} 	& 0
	\end{bmatrix}
	\begin{bmatrix}
	{\hat h}^{({L})} \\
	{\hat \phi}
	\end{bmatrix}, \label{eq:gravity41b}\\
\delta%_{\theta_{6}}
	\begin{bmatrix}
	{\hat h}^{({\rm tr})} \\
	{\hat h}_{5}^{({\rm L})}
	\end{bmatrix}
&= 	\theta_{6}
	\begin{bmatrix}
	0 			& \partial_{z} + 3A' \\
	\partial_{z} 	& 0
	\end{bmatrix}
	\begin{bmatrix}
	{\hat h}^{({\rm tr})} \\
	{\hat h}_{5}^{({\rm L})}
	\end{bmatrix}, \label{eq:gravity41c}\\
\delta%_{\theta_{7}}
	\begin{bmatrix}
	{\hat h}_{5}^{({\rm L})} \\
	{\hat \phi}
	\end{bmatrix}
&= 	\theta_{7}
	\begin{bmatrix}
	0 				& \partial_{z} + 2A' \\
	-(\partial_{z} + A') 	& 0
	\end{bmatrix}
	\begin{bmatrix}
	{\hat h}_{5}^{({\rm L})} \\
	{\hat \phi}
	\end{bmatrix}. \label{eq:gravity41d}
\end{align}
\end{subequations}
It is interesting to note that the $\theta_{7}$-symmetry is generated by ${\bar Q}_{1}$ in \eqref{eq:gravity15b}.

(4)
The residual symmetries \eqref{eq:gravity38a}, \eqref{eq:gravity38b}, and \eqref{eq:gravity38c} are spacetime symmetries between a spin-2 and a spin-1 particles, a spin-2 and a spin-1 particles and a spin-2 and a spin-0 particles, respectively.
Because of the same reason in the Abelian gauge theory with extra compact dimensions in Section \ref{sec:Abelian}, it does not contradict with the Coleman-Mandula theorem.

%%%%% SECTION 4 %%%%%
\section{One Graviton Exchange Amplitude} \label{sec:amplitude}
In this section we will calculate the lowest tree-level amplitude for the process of a gravitational interaction between two (conserved) external energy-momentum tensors residing on the 3-branes by using the previously derived $R_{\xi}$ gauge, and show that the result is actually independent of the gauge parameter $\xi$.

\begin{figure*}[t]
	\begin{align}
	i{\cal M}
	&= 	\sum_{n=0}^{\infty}
		\begin{array}{c}
		\includegraphics{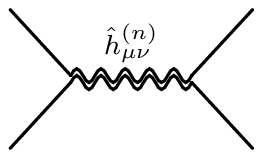}
		\end{array}
		+
		\sum_{n=0}^{\infty}
		\begin{array}{c}
		\includegraphics{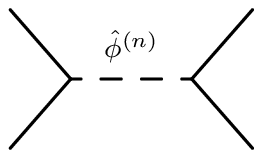}
		\end{array} \nonumber
	\end{align}
\caption{One graviton exchange amplitude.}
\label{fig:amp}
\end{figure*}

The coupling of the fluctuations ${\bar h}_{\mu\nu}$ to the energy-momentum tensor $T_{\mu\nu}$ is
\begin{align}
& 	- \frac{1}{2}{\bar h}_{\mu\nu}T^{\mu\nu}\delta(z-z_{i})
= 	- \frac{1}{2}\left(h_{\mu\nu} - \frac{1}{2}\eta_{\mu\nu}\phi\right)T^{\mu\nu}\delta(z-z_{i}) \nonumber\\
&= 	\left(
	- \frac{1}{\sqrt{2}M_{\rm Pl}}{\hat h}_{\mu\nu}T^{\mu\nu}
	+ \frac{1}{2\sqrt{3}M_{\rm Pl}}{\hat \phi}T
	\right)\delta(z-z_{i}), \label{eq:amplitude01}
\end{align}
where $z_{1,2} = 0$ or $z_{c}$.
The last equality follows from the canonically normalized fields ${\hat h}_{\mu\nu} = (M_{\rm Pl}/\sqrt{2})h_{\mu\nu}$ and ${\hat \phi} = (\sqrt{3}M_{\rm Pl}/2)\phi$.
The scattering amplitude for our process is given by (see FIG.\ref{fig:amp})
\begin{align}
i{\cal M}
&= 	\sum_{n=0}^{\infty}
	\left(\frac{-if^{(n)}(z_{1})}{\sqrt{2}{M_{\rm Pl}}}{\tilde T}^{(1)\mu\nu}\right)
	{\tilde \Delta}^{{\hat h}_{\mu\nu}}_{\{\mu\nu,n\}; \{\rho\sigma,n\}}(k) \nonumber\\
&	\hspace{1em}
	\times
	\left(\frac{-if^{(n)}(z_{2})}{\sqrt{2}{M_{\rm Pl}}}{\tilde T}^{(2)\rho\sigma}\right) \nonumber\\
& 	\hspace{1em}
	+ \sum_{n=0}^{\infty}
	\left(\frac{ik^{(n)}(z_{1})}{2\sqrt{3}{M_{\rm Pl}}}{\tilde T}^{(1)}\right)
	{\tilde \Delta}^{\hat \phi}_{\{n\}; \{n\}}(k)
	\left(\frac{ik^{(n)}(z_{2})}{2\sqrt{3}{M_{\rm Pl}}}{\tilde T}^{(2)}\right), \label{eq:amplitude02}
\end{align}
Note that, as shown in Appendix \ref{appendix:propagator}, the massive graviton propagator can be separated into the transverse-traceless part $\Delta^{(\rm TT)}$, the transverse gauge freedom part $\Delta^{(\rm T)}$, the longitudinal gauge freedom part $\Delta^{(\rm L)}$ and the trace part $\Delta^{(\rm tr)}$.
Since $\Delta^{(\rm T)}$ and $\Delta^{(\rm L)}$ do not couple to the conserved energy momentum tensor, the cancellation of the $\xi$-dependence for $n>0$ must take place between the trace part of $h_{\mu\nu}$ and the would-be NG boson $\phi$.
Its cancellation certainly occurs in a bit surprising way:
\begin{widetext}
\begin{align}
&	\hspace{1em}
	- \frac{f^{(n)}(z_{1})f^{(n)}(z_{2})}{2M_{\rm Pl}^{2}}
	{\tilde T}^{(1)\mu\nu}
	{\tilde \Delta}^{({\rm tr})}_{\{\mu\nu,n\};\{\rho\sigma,n\}}
	{\tilde T}^{(2)\rho\sigma}
	- \frac{k^{(n)}(z_{1})k^{(n)}(z_{2})}{12M_{\rm Pl}^{2}}
	{\tilde T}^{(1)}
	{\tilde \Delta}^{\hat \phi}_{\{n\}; \{n\}}
	{\tilde T}^{(2)} \nonumber\\
&= 	- \frac{{\tilde T}^{(1)}{\tilde T}^{(2)}}{12M_{\rm Pl}^{2}}
	\frac{i}{- k^{2} - (3\xi - 2)m_{n}^{2}}
	\bigl[- f^{(n)}(z_{1})f^{(n)}(z_{2}) + k^{(n)}(z_{1})k^{(n)}(z_{2})\bigr]
= 	0, \label{eq:amplitude03}
\end{align}
\end{widetext}
where the last equality follows from the identity \eqref{eq:gravity14} and the boundary conditions \eqref{eq:gravity12}.
Thus the resulting amplitude does not depend on a gauge choice thanks to the supersymmetries between the eigenfunctions.
We finally arrive at the expression
\begin{align}
i{\cal M}
&= 	\frac{-i}{2M_{\rm Pl}^{2}}\frac{f^{(0)}(z_{1})f^{(0)}(z_{2})}{-k^{2}}
	\left\{{\tilde T}^{(1)\mu\nu}{\tilde T}^{(2)}_{\mu\nu}
	- \frac{1}{2}{\tilde T}^{(1)}{\tilde T}^{(2)}\right\} \nonumber\\
&	+ \frac{-i}{12M_{\rm Pl}^{2}}\frac{k^{(0)}(z_{1})k^{(0)}(z_{2})}{-k^{2}}
	{\tilde T}^{(1)}{\tilde T}^{(2)} \nonumber\\
& 	+ \frac{-i}{2M_{\rm Pl}^{2}}
	\sum_{n=1}^{\infty}\frac{f^{(n)}(z_{1})f^{(n)}(z_{2})}{-k^{2}-m_{n}^{2}}
	\left\{{\tilde T}^{(1)\mu\nu}{\tilde T}^{(2)}_{\mu\nu}
	- \frac{1}{3}{\tilde T}^{(1)}{\tilde T}^{(2)}\right\}. \label{eq:amplitude04}
\end{align}
It is worth mentioning that in our normalization the zero-mode functions are
\begin{equation}
f^{(0)}(z) = 1
\qquad\text{and}\qquad
k^{(0)}(z) = {\rm e}^{-\pi kR}(1 + kz)^{2}, \label{eq:amplitude05}
\end{equation}
so that if we set $z_{1} = 0$ and $z_{2} = z_{c}$, $f^{(0)}(0)f^{(0)}(z_{c})$ and $k^{(0)}(0)k^{(0)}(z_{c})$ give the same value and are equal to one.
This means that for the process of a gravitational interaction between two energy-momentum tensors sitting on the UV and IR branes, the massless limit $m_{n}\to0$ is  continuous even in the lowest tree-level approximation.
Thus, in this special case it may be possible to evade the vDVZ-discontinuity without invoking (anti-)de Sitter backgrounds \cite{Porrati:2001,KMP:2001,KKR:2001,Neupane:2001dv} nor nonperturbative effects \cite{Vainshtein:1972sx,DDGV:2002}.
On the other hand, if we set $z_{1}=z_{2}=0$ or $z_{1}=z_{2}=z_{c}$, the massless radion's contribution could not coincide with the massless graviton's one.
In these cases the massless limit is not continuous and hence the vDVZ-discontinuity appears in the amplitude.

\subsection{vDVZ-Discontinuity from Supersymmetric Point of View}
As already mentioned, it is known that there exists a discontinuity between the massless graviton and the massive graviton theory in the lowest tree-level approximation with Minkowski background \cite{Iwasaki:1970,vDV:1970,Zakharov:1970}.
As summarized in Appendix \ref{appendix:vDVZ}, the discontinuity is due to the fact that the helicity 0 state dose not decouple to the trace of a conserved energy-momentum tensor even in the massless limit.

In the $R_{\xi}$ gauge the fictitious scalar fields propagate on the branes, however, as mentioned before, we have the identity $f^{(n)}(z_{i}) = - k^{(n)}(z_{i})$ for $n>0$ on the branes, which is nothing but the consequence of the supersymmetries in the mass spectrum, such that the massive scalars' contributions can be converted into the massive gravitons' one.
However, the zero modes $f^{(0)}(z_{i})$ and $k^{(0)}(z_{i})$ do not mutually relate so that the massless graviton's contribution differs from the massive graviton's one, and then there arises the discontinuity.

%%%%% SECTION 5 %%%%%
\section{Summary} \label{sec:summary}
In this paper we have studied higher-dimensional gauge/gravity theories with compact extra dimension(s).
In Section \ref{sec:Abelian} we have investigated the pure Abelian gauge theory with extra compact $D$-dimensions in the $R_{\xi}$ gauge, and found that the gauge-fixed action possesses the residual global symmetries between unphysical component of the gauge fields and the would-be scalar NG bosons.
The $\xi$-dependent mass-matrix of this multiplet is exactly the super-Hamiltonian and the residual global symmetry is generated by the supercharge found in the analysis of the mass spectrum.

In Section \ref{sec:gravity} we have extended the analysis to five-dimensional pure gravity to quadratic order as perturbation to the Randall-Sundrum background.
We have established the one-parameter family of gauge choices for 5d gravity analogous to the $R_{\xi}$ gauge in spontaneously broken gauge theories.
In this $R_{\xi}$ gauge we have shown that the gauge-fixed action exhibits the three independent residual global symmetries.
As in the pure Abelian gauge theory with extra dimensions, it is the symmetries between unphysical component of the graviton field and the would-be vector and scalar NG bosons.
However, as opposed to the pure Abelian gauge theories with extra dimensions, one of these residual symmetry transformations cannot be expressed in terms of the first-order differential supercharges but the second-order differential supercharge.

In Section \ref{sec:amplitude}, to see the validity of our $R_{\xi}$ gauge we have computed the one graviton exchange amplitude in the lowest tree-level approximation, and confirmed that it is indeed $\xi$-independent.
Using this result we also found that for the process of a gravitational interaction between two conserved energy-momentum tensors residing on the UV and IR branes, the famous vDVZ-discontinuity does not arise even in the lowest tree-level approximation.

%%%%% Acknowledgments %%%%%%%%%%
\section*{ACKNOWLEDGMENTS}
C.S.L. and M.S. are supported in part by the Grant-in-Aid for Scientific Research (No.18204024 and No.18540275) by the Japanese Ministry of Education, Science, Sports and Culture.
The authors would like to thank K.Ghoroku, N.Maru, K.Sato, H.Sonoda, M.Tachibana and K.Takenaga for valuable discussions.

\appendix
%%%%% APPENDIX A %%%%%
\section{vDVZ-Discontinuity} \label{appendix:vDVZ}
In the massive 4d graviton theory with the Fierz-Pauli mass term, the extra scalar degree of freedom corresponding to the helicity 0 state does not decouple to a trace of a stress-energy tensor even in the limit $m\to0$.
This is the famous van Dam-Veltman-Zakharov (vDVZ) no-go theorem, that is, the helicity 0 degree of freedom causes the discontinuity between the massless graviton theory and the massless limit of the massive graviton theory.

In this Appendix we briefly review a massless and massive graviton theory in four-dimensional Minkowski background and discuss how the extra helicity 0 degree of freedom comes in to the propagator and couples to the conserved energy-momentum tensor.

%%%%% APPENDIX A.1 %%%%%
\subsection{Massless 4d Graviton} \label{appendix:massless}
We start out with the action
\begin{align}
S
= 	\left.M^{2}\int{\rm d}^{4}x\sqrt{-g}R\right|_{O(h^{2})}
= 	M^{2}\int{\rm d}^{4}x~
	\frac{1}{4}h^{\mu\nu}K_{\mu\nu; \rho\sigma}h^{\rho\sigma}, \label{eq:massless01}
\end{align}
where
\begin{align}
K_{\mu\nu;\rho\sigma}
&=	- \frac{1}{2}
	\left(\eta_{\mu\rho}\partial_{\nu}\partial_{\sigma}
	+ \eta_{\mu\sigma}\partial_{\nu}\partial_{\rho}
	+ \eta_{\nu\rho}\partial_{\mu}\partial_{\sigma}
	+ \eta_{\nu\sigma}\partial_{\mu}\partial_{\rho}\right) \nonumber\\
&	\hspace{1em}
	+ \eta_{\mu\nu}\partial_{\rho}\partial_{\sigma}
	+ \eta_{\rho\sigma}\partial_{\mu}\partial_{\nu} \nonumber\\
&	\hspace{1em}
	+ \frac{1}{2}\left(\eta_{\mu\rho}\eta_{\nu\sigma}
	+ \eta_{\mu\sigma}\eta_{\nu\rho}
	- 2\eta_{\mu\nu}\eta_{\rho\sigma}\right)\Box. \label{eq:massless02}
\end{align}
$M$ is the fundamental mass scale which is determined in terms of the Newton constant $G_{N}$ later.
Our flat metric $\eta_{\mu\nu}$ has the signature $(-, +, +, +)$.
The action is invariant under the linearized general coordinate transformations:
\begin{align}
h_{\mu\nu}
\mapsto
{\hat h}_{\mu\nu}
= h_{\mu\nu} - \partial_{\mu}\xi_{\nu} - \partial_{\nu}\xi_{\mu}, \label{eq:massless03}
\end{align}
where $\xi_{\mu}$ are four arbitrary functions.

The equation of motion (EOM) is
\begin{align}
0
&= 	\partial_{\mu}\partial^{\rho}h_{\rho\nu}
	+ \partial_{\nu}\partial^{\rho}h_{\rho\mu}
	- \eta_{\mu\nu}\partial_{\rho}\partial_{\sigma}h^{\rho\sigma}
	- \partial_{\mu}\partial_{\nu}h \nonumber\\
& 	\hspace{1em}
	- \Box(h_{\mu\nu} - \eta_{\mu\nu}h), \label{eq:massless04}
\end{align}
where $h := \eta^{\mu\nu}h_{\mu\nu}$.
We apply $\eta^{\mu\nu}$ to Eq.\eqref{eq:massless04} with the result
\begin{align}
0
&= 	\partial^{\nu}\partial^{\rho}h_{\rho\nu}
	+ \partial^{\mu}\partial^{\rho}h_{\rho\mu}
	- 4\partial_{\rho}\partial_{\sigma}h^{\rho\sigma}
	- \Box h
	- \Box(h - 4h) \nonumber\\
&= 	- 2\partial_{\mu}\partial_{\nu}h^{\mu\nu} + 2\Box h. \label{eq:massless05}
\end{align} 
Eq.\eqref{eq:massless04} then reduces to
\begin{align}
0
= 	\partial_{\mu}\partial^{\rho}h_{\rho\nu}
	+ \partial_{\nu}\partial^{\rho}h_{\rho\mu}
	- \partial_{\mu}\partial_{\nu}h
	- \Box h_{\mu\nu}. \label{eq:massless06}
\end{align}
To see the physical degrees of freedom, we impose the harmonic gauge conditions:
\begin{align}
- \partial^{\mu}{\hat h}_{\mu\nu} + \frac{1}{2}\partial_{\nu}{\hat h} = 0. \label{eq:massless07}
\end{align}
Notice that Eq.\eqref{eq:massless07} is the linearized version of $\partial_{\mu}\left(\sqrt{- g}g^{\mu\nu}\right)$.
This is one of the crucial keys to construct a gauge-fixing term for 5d gravity; see Section \ref{subsec:gravity_gauge-fix}.

The conditions \eqref{eq:massless07} are implemented by choosing the gauge parameters $\xi_{\mu}$ as solutions to the following differential equations:
\begin{align}
\Box \xi_{\mu} = \partial^{\nu}h_{\nu\mu} - \frac{1}{2}\partial_{\mu}h. \label{eq:massless08}
\end{align}
EOM now becomes
\begin{align}
\Box {\hat h}_{\mu\nu} = 0. \label{eq:massless09}
\end{align}
Note that there still remains gauge freedom generated by $\xi_{\mu}$ which satisfy $\Box \xi_{\mu} = 0$. 
If we decompose on-shell parameters $\xi_{\mu}$ into transverse part (3 DOF) and longitudinal part (1 DOF), $\xi_{\mu} = \xi_{\mu}^{(T)} + \xi_{\mu}^{(L)}$, the combination $\partial_{\mu}\xi_{\nu} + \partial_{\nu}\xi_{\mu}$ becomes
\begin{align}
\partial_{\mu}\xi_{\nu} + \partial_{\nu}\xi_{\mu}
&= 	\underbrace{\partial_{\mu}\xi_{\nu}^{(T)}
	+ \partial_{\nu}\xi_{\mu}^{(T)}}_{\textrm{transverse-traceless}} \nonumber\\
&	\hspace{1em}
	+ \underbrace{\partial_{\mu}\xi_{\nu}^{(L)}
	+ \partial_{\nu}\xi_{\mu}^{(L)}
	- \frac{1}{2}\eta_{\mu\nu}\partial\xi^{L}}_{\textrm{traceless but not transverse}} \nonumber\\
&	\hspace{1em}
	+ \underbrace{\frac{1}{2}\eta_{\mu\nu}\partial\xi^{(L)}}_{\textrm{trace piece}}.
	\label{eq:massless10}
\end{align}
where $\partial\xi^{(L)} := \partial^{\mu}\xi_{\mu}^{(L)}$. 
When we choose $\xi_{\mu}^{(L)}$ as a solution to the equation $2\partial\xi^{(L)} = {\hat h}$, the fluctuations become traceless and then, because of the harmonic gauge conditions, automatically satisfy transverse conditions $\partial^{\mu}{\hat h}_{\mu\nu} = 0$.
Now that ${\hat h}_{\mu\nu}$ is transverse-traceless and has five degrees of freedom, however, 3 out of 5 DOF can be eliminated by $\xi_{\mu}^{(T)}$.
Thus we see that $h_{\mu\nu}$ has two physical degrees of freedom. This is the helicity degrees of freedom for massless graviton.

Next gauge-fix the action. By adding a gauge-fixing term ${\cal L}_{\rm GF} = M^{2}(- 1/2)(-\partial_{\mu}h^{\mu\nu} + (1/2)\partial^{\nu}h)^{2}$, we can eliminate the first two lines in Eq.\eqref{eq:massless02} so that the action becomes
\begin{align}
S
&= 	M^{2}\!\!\int{\rm d}^{4}x
	\frac{1}{4}h^{\mu\nu}
	\left[\frac{1}{2}
	\left(\eta_{\mu\rho}\eta_{\nu\sigma}
	+ \eta_{\mu\sigma}\eta_{\nu\rho}
	- \eta_{\mu\nu}\eta_{\rho\sigma}\right)
	\Box\right]h^{\rho\sigma} \nonumber\\
&= 	\int{\rm d}^{4}x
	\frac{1}{2}{\hat h}^{\mu\nu}
	\left[\frac{1}{2}
	\left(\eta_{\mu\rho}\eta_{\nu\sigma}
	+ \eta_{\mu\sigma}\eta_{\nu\rho}
	- \eta_{\mu\nu}\eta_{\rho\sigma}\right)
	\Box\right]{\hat h}^{\rho\sigma}\label{eq:massless11}
\end{align}
where we have rescaled the field $h^{\mu\nu} \mapsto {\hat h}^{\mu\nu} = (M/\sqrt{2})h^{\mu\nu}$ to be the canonical action.

Noting that the symmetric unit bitensor is expressed as $I_{\mu\nu; \rho\sigma} = (1/2)(\eta_{\mu\rho}\eta_{\nu\sigma} + \eta_{\mu\sigma}\eta_{\nu\rho})$, it is easy to check that the inverse of a bitensor $(1/2)\left(\eta_{\mu\rho}\eta_{\nu\sigma} + \eta_{\mu\sigma}\eta_{\nu\rho} - \eta_{\mu\nu}\eta_{\rho\sigma}\right)$ is actually itself.
Thus, in the harmonic gauge, the free propagator for massless graviton is given by
\begin{align}
{\tilde \Delta}_{\mu\nu; \rho\sigma}(k)
= 	\frac{i}{2}
	\frac{\eta_{\mu\rho}\eta_{\nu\sigma}
	+ \eta_{\mu\sigma}\eta_{\nu\rho}
	- \eta_{\mu\nu}\eta_{\rho\sigma}}
	{- k^{2}}. \label{eq:massless12}
\end{align}

Next determine the mass scale $M$ to give the Newton constant $G_{N}$ by computing the static gravitational potential between two masses, $\mu_{1}$ and $\mu_{2}$.
The corresponding amplitude is
\begin{align}
-i{\tilde V}({\vec k})
&= 	\lim_{k^{0}\to0}
	\frac{i\mu_{1}}{\sqrt{2}M} \cdot
	\frac{i}{2}\frac{\eta_{00}\eta_{00} + \eta_{00}\eta_{00} - \eta_{00}\eta_{00}}{- k^{2}} \cdot
	\frac{i\mu_{2}}{\sqrt{2}M} \nonumber\\
&= 	\frac{\mu_{1}\mu_{2}}{4M^{2}}\frac{i}{{\vec k}^{2}}, \label{eq:massless13}
\end{align}
which gives the static potential $V(r) = - (\mu_{1}\mu_{2}/4M^{2})(1/4\pi r^{2})$ so that it should be chosen that $M = 1/\sqrt{16\pi G_{N}}$.

Next examine the tensor structure of the residue of the propagator \eqref{eq:massless12}.
In $p^{\mu} = {}^{t}(p, 0, 0, p)$ frame, we have two independent polarization tensors
\begin{align}
\epsilon_{\mu\nu}^{(\pm 2)}
= 	\frac{1}{2}
	\begin{bmatrix}
	0 	& 0 		& 0 		& 0 \\
	0 	& 1 		& \pm i 	& 0 \\
	0 	& \pm i 	& -1 		& 0 \\
	0 	& 0 		& 0 		& 0
	\end{bmatrix}, \label{eq:massless14}
\end{align}
where we have normalized them with respect to the inner product
\begin{align}
(\epsilon_{\mu\nu}^{(i)}, \epsilon_{\rho\sigma}^{(j)})
:= \eta^{\mu\rho}\eta^{\nu\sigma}\epsilon_{\mu\nu}^{(i)}\epsilon_{\rho\sigma}^{(j)\ast},
\qquad
i, j = \pm 2. \label{eq:massless15}
\end{align}
As explicitly constructed in the next subsection, \eqref{eq:massless14} is given by the tensor product of the polarization vectors corresponding to the helicity $\pm 1$ states as in a manner similar to the addition of angular momentum in quantum mechanics.

Using Eq.\eqref{eq:massless14} one can verify the following relation:
\begin{align}
& 	\sum_{i = \pm 2} \epsilon_{\mu\nu}^{(i)}\epsilon_{\rho\sigma}^{(i)\ast} \nonumber\\
&= 	\begin{cases}
	\frac{1}{2}\left(\eta_{\mu\rho}\eta_{\nu\sigma}
	+ \eta_{\mu\sigma}\eta_{\nu\rho}
	- \eta_{\mu\nu}\eta_{\rho\sigma}\right) 	& \mu, \nu, \rho, \sigma = 1, 2, \\
	0 									& \text{otherwise}. \label{eq:massless16}
	\end{cases}
\end{align}
Indeed, $\sum_{i = \pm 2}\epsilon_{11}^{(i)}\epsilon_{11}^{(i)\ast} = 1/4 + 1/4 = 1/2$ while $(1/2)(\eta_{11}\eta_{11} + \eta_{11}\eta_{11} - \eta_{11}\eta_{11}) = 1/2$, $\sum_{i = \pm 2}\epsilon_{12}^{(i)}\epsilon_{12}^{(i)\ast} = 1/4 + 1/4 = 1/2$ while $(1/2)(\eta_{11}\eta_{22} + \eta_{12}\eta_{21} - \eta_{12}\eta_{12}) = 1/2$, and so on.
From this, we see that the residue of the propagator \eqref{eq:massless12} has the correct polarization structure.

%%%%% APPENDIX A.2 %%%%%
\subsection{Massive 4d Graviton} \label{appendix:massive_graviton}
Next consider the massive graviton theory by adding a Fierz-Pauli \cite{FP:1939} mass term $M_{m}^{2}(-1/4)m^{2}(h_{\mu\nu}h^{\mu\nu} - h^{2})$:
\index{Fierz-Pauli mass term}
\begin{align}
S
= 	M_{m}^{2}\int{\rm d}^{4}x~
	\frac{1}{4}h^{\mu\nu}K^{m}_{\mu\nu; \rho\sigma}h^{\rho\sigma}, \label{eq:massive01}
\end{align}
where
\begin{align}
K^{m}_{\mu\nu;\rho\sigma}
&=	- \frac{1}{2}
	\left(\eta_{\mu\rho}\partial_{\nu}\partial_{\sigma}
	+ \eta_{\mu\sigma}\partial_{\nu}\partial_{\rho}
	+ \eta_{\nu\rho}\partial_{\mu}\partial_{\sigma}
	+ \eta_{\nu\sigma}\partial_{\mu}\partial_{\rho}\right) \nonumber\\
&	\hspace{1em}
	+ \eta_{\mu\nu}\partial_{\rho}\partial_{\sigma}
	+ \eta_{\rho\sigma}\partial_{\mu}\partial_{\nu} \nonumber\\
&	\hspace{1em}
	+ \frac{1}{2}\left(\eta_{\mu\rho}\eta_{\nu\sigma}
	+ \eta_{\mu\sigma}\eta_{\nu\rho}
	- 2\eta_{\mu\nu}\eta_{\rho\sigma}\right)
	(\Box - m^{2}). \label{eq:massive02}
\end{align}
$M_{m}$ is the corresponding mass scale for the massive graviton theory.
The resulting EOM is
\begin{align}
0
&= 	\partial_{\mu}\partial^{\rho}h_{\rho\nu}
	+ \partial_{\nu}\partial^{\rho}h_{\rho\mu}
	- \eta_{\mu\nu}\partial_{\rho}\partial_{\sigma}h^{\rho\sigma}
	- \partial_{\mu}\partial_{\nu}h \nonumber\\
&	\hspace{1em}
	+ (- \Box + m^{2})(h_{\mu\nu} - \eta_{\mu\nu}h). \label{eq:massive03}
\end{align}
By contracting and taking the trace of Eq.\eqref{eq:massive03} we have
\begin{align}
0
&= 	m^{2}\left(\eta_{\mu\rho}\partial_{\sigma} - \eta_{\rho\sigma}\partial_{\mu}\right)
	h^{\rho\sigma} \nonumber\\
\text{and} \quad
0
&= 	(\partial_{\rho}\partial_{\sigma} - \eta_{\rho\sigma}\Box)h^{\rho\sigma}
	+ \frac{3}{2}m^{2}h. \label{eq:massive04}
\end{align}
Since $m\neq0$, these equations can farther be reduced to the following constraints:
\begin{align}
h = 0
\quad\textrm{and}\quad
\partial_{\mu}h^{\mu\nu} = 0, \label{eq:massive05}
\end{align}
which provide $1 + 4 = 5$ conditions, so that the EOM now becomes
\begin{align}
(- \Box + m^{2})h^{\mu\nu} = 0. \label{eq:massive06}
\end{align}
Now it is obvious that 5 out of 10 DOF are eliminated by the traceless-transverse conditions \eqref{eq:massive05} so that $h^{\mu\nu}$ has only $10 - 5 = 5$ DOF, corresponding to the helicity degrees of freedom of a massive spin-2 particle.
This is why the Fierz-Pauli mass term $-(1/2)m^{2}(h^{\mu\nu}h_{\mu\nu} - h^{2})$ gives the graviton a mass in a natural way.

The massive graviton propagator $\Delta^{\mu\nu; \rho\sigma}(x-y)$ is defined as the Green function inverse to the differential operator $K^{m}_{\mu\nu; \kappa\lambda}$:
\begin{align}
K^{m}_{\mu\nu; \kappa\lambda}\Delta^{\mu\nu; \rho\sigma}(x-y)
= 	\frac{i}{2}
	({\delta^{\rho}}_{\kappa}{\delta^{\sigma}}_{\lambda}
	+ {\delta^{\rho}}_{\lambda}{\delta^{\sigma}}_{\kappa})
	\delta^{(4)}(x-y). \label{eq:massive07}
\end{align}
After tedious calculations we get
\begin{align}
{\tilde \Delta}_{\mu\nu; \rho\sigma}(k)
&= 	\frac{i}{-k^{2}-m^{2}}
	\biggl[\frac{1}{2}\left(\eta_{\mu\rho} + \frac{k_{\mu}k_{\rho}}{m^{2}}\right)
	\left(\eta_{\nu\sigma} + \frac{k_{\nu}k_{\sigma}}{m^{2}}\right) \nonumber\\
&	\hspace{1em}
	+ \frac{1}{2}\left(\eta_{\mu\sigma} + \frac{k_{\mu}k_{\sigma}}{m^{2}}\right)
	\left(\eta_{\nu\rho} + \frac{k_{\nu}k_{\rho}}{m^{2}}\right) \nonumber\\
&	\hspace{1em}
	- \frac{1}{3}\left(\eta_{\mu\nu} + \frac{k_{\mu}k_{\nu}}{m^{2}}\right)
	\left(\eta_{\rho\sigma} + \frac{k_{\rho}k_{\sigma}}{m^{2}}\right)\biggr],
	\label{eq:massive08}
\end{align}
in the momentum space.

Let us compute the static gravitational potential via exchange of a massive graviton between two masses, $\mu_{1}$ and $\mu_{2}$.
The corresponding amplitude is
\begin{align}
-i{\tilde V}_{m}({\vec k})
&= 	\lim_{k^{0}\to0}
	\left(\frac{i\mu_{1}}{\sqrt{2}M_{m}}\right)\left(\frac{i\mu_{2}}{\sqrt{2}M_{m}}\right) \nonumber\\
&	\hspace{1em}
	\times
	\frac{i}{2}
	\frac{\eta_{00}\eta_{00} + \eta_{00}\eta_{00} - (2/3)\eta_{00}\eta_{00}
	+ k^{0}~\textrm{terms}}{- k^{2} - m^{2}} \nonumber\\
&= 	\frac{4}{3}\cdot\frac{\mu_{1}\mu_{2}}{4M_{m}^{2}}\frac{i}{{\vec k}^{2} + m^{2}},
	\label{eq:massive09}
\end{align}
which gives the static potential $V_{m}(r) = - (4/3)(\mu_{1}\mu_{2}/4M_{m}^{2})({\rm e}^{-mr}/4\pi r^{2})$.
If we require $V_{m}(r)$ with an extremely tiny but non-vanishing mass $m$ gives the same result as that of the exchange of a massless graviton, the Newton constant $G_{N}$ should be related as $G_{m} = (3/4)G_{N}$, where $G_{m} := 1/(16\pi M_{m}^{2})$ is the coupling constant for the massive graviton theory.
This result that the coupling constant for the massive graviton theory $G_{m}$ is 25\% smaller than that of the massless graviton theory $G_{N}$ yields the different prediction for the bending of light (photon, whose energy-momentum tensor is traceless) passing near the Sun, whose experimental test (measurement of deflection angle) has been performed with the accuracy $\sim$1\% to the prediction for the standard general relativity, while the massive graviton theory's prediction is about 25\% smaller than that of the massless graviton theory, which contradicts with the measurement.
This difference is caused by the helicity 0 degree of freedom which couples to the trace of an energy-momentum tensor even in the massless limit.

To see this let us finally examine the polarization structure of the residue of the massive graviton propagator.
To this end, we move on to the rest frame for the massive graviton: $p^{\mu} = {}^{t}(m, 0, 0, 0)$.
In this frame with the Cartesian coordinate basis, the five independent polarization tensors are given by
\begin{subequations}
\begin{align}
\epsilon^{(\pm 2)}_{\mu\nu}
&= 	\epsilon^{(\pm 1)}_{\mu}\otimes\epsilon^{(\pm 1)}_{\nu}
= 	\frac{1}{2}
		\begin{bmatrix}
		0 	& 0 		& 0 		& 0 \\
		0 	& 1 		& \pm i 	& 0 \\
		0 	& \pm i 	& -1 		& 0 \\
		0 	& 0 		& 0 		& 0
		\end{bmatrix}, \label{eq:massive10a}\\
\epsilon^{(\pm 1)}_{\mu\nu}
&= 	\frac{1}{\sqrt{2}}
	\bigl(
	\epsilon^{(\pm 1)}_{\mu}\otimes\epsilon^{(0)}_{\nu}
	+ \epsilon^{(0)}_{\mu}\otimes\epsilon^{(\pm 1)}_{\nu}
	\bigr) \nonumber\\
&= 	\frac{1}{2}
		\begin{bmatrix}
		0 	& 0 		& 0 		& 0 \\
		0 	& 0 		& 0		& \pm 1 \\
		0 	& 0		& 0 		& i \\
		0 	& \pm 1	& i		& 0
		\end{bmatrix}, \label{eq:massive10b}\\
\epsilon^{(0)}_{\mu\nu}
&= 	\frac{1}{\sqrt{6}}
	\bigl(
	\epsilon^{(\pm 1)}_{\mu}\otimes\epsilon^{(\mp 1)}_{\nu}
	+ \epsilon^{(\mp 1)}_{\mu}\otimes\epsilon^{(\pm 1)}_{\nu}
	+ 2\epsilon^{(0)}_{\mu}\otimes\epsilon^{(0)}_{\nu}
	\bigr) \nonumber\\
&= 	\sqrt{\frac{2}{3}}
		\begin{bmatrix}
		0 	& 0 		& 0 		& 0 \\
		0 	& -1/2 	& 0 		& 0 \\
		0 	& 0  		& -1/2 	& 0 \\
		0 	& 0 		& 0		& 1
		\end{bmatrix}. \label{eq:massive10c}
\end{align}
\end{subequations}
where $\epsilon^{(\pm 1)}_{\mu}$ and $\epsilon^{(0)}_{\mu}$ are the polarization vectors for a massive vector boson with mass $m$:
\begin{align}
\epsilon^{(\pm 1)}_{\mu}
= 	\frac{1}{\sqrt{2}}
	\begin{bmatrix}
	0 \\
	\pm 1 \\
	i \\
	0
	\end{bmatrix}, \quad
\epsilon^{(0)}_{\mu}
= 	\begin{bmatrix}
	0 \\
	0 \\
	0 \\
	1
	\end{bmatrix},
\end{align}
and $\otimes$ denotes the tensor product defined as
\begin{align}
 	\begin{bmatrix}
	a_{0} \\
	a_{1} \\
	a_{2} \\
	a_{3}
	\end{bmatrix}
	\otimes
	\begin{bmatrix}
	b_{0} \\
	b_{1} \\
	b_{2} \\
	b_{3}
	\end{bmatrix}
:= 	\begin{bmatrix}
	a_{0}b_{0} 	& a_{0}b_{1} 	& a_{0}b_{2} 	& a_{0}b_{3} \\
	a_{1}b_{0} 	& a_{1}b_{1} 	& a_{1}b_{2} 	& a_{1}b_{3} \\
	a_{2}b_{0} 	& a_{2}b_{1} 	& a_{2}b_{2} 	& a_{2}b_{3} \\
	a_{3}b_{0} 	& a_{3}b_{1} 	& a_{3}b_{2} 	& a_{3}b_{3}
	\end{bmatrix}, \quad
	a_{i}, b_{j}\in\mathbb{C}.
\end{align}
One can easily verify that
\begin{align}
& 	\sum_{i = \pm 2, \pm1, 0}\epsilon^{(i)}_{\mu\nu}\epsilon^{(i)\ast}_{\rho\sigma} \nonumber\\
&= 	\left\{\begin{array}{ll}
	\frac{1}{2}\left(\eta_{\mu\rho}\eta_{\nu\sigma}
	+ \eta_{\mu\sigma}\eta_{\nu\rho}
	- \frac{2}{3}\eta_{\mu\nu}\eta_{\rho\sigma}\right)
	& \mu, \nu, \rho, \sigma = 1, 2, 3,\\
	0
	& \text{otherwise}
	\end{array}\right. \label{eq:massive12}
\end{align}
Indeed, $\sum\epsilon^{(i)}_{11}\epsilon^{(i)\ast}_{11} = 1/4 + 1/4 + 0 + 0 + (2/3)\cdot (1/4) = 2/3$ while $(1/2)(\eta_{11}\eta_{11} + \eta_{11}\eta_{11} - (2/3)\eta_{11}\eta_{11}) = 2/3$, $\sum\epsilon^{(i)}_{12}\epsilon^{(i)\ast}_{12} = 1/4 + 1/4 + 0 + 0 + 0 = 1/2$ while $(1/2)(\eta_{11}\eta_{22} + \eta_{12}\eta_{21} - (2/3)\eta_{12}\eta_{12}) = 1/2$, and so on.

It should be noted that Eq.\eqref{eq:massive12} is different from Eq.\eqref{eq:massless15} even if $\mu, \nu, \rho, \sigma = 1, 2$.
This ``2/3-discontinuity'' is due to the existence of the helicity 0 state $\epsilon^{(0)}_{\mu\nu}$, which was first pointed out in 1970 independently by Iwasaki \cite{Iwasaki:1970}, and by van Dam and Veltman \cite{vDV:1970}, and by Zakharov \cite{Zakharov:1970}.

%%%%% APPENDIX B %%%%%
\section{Propagators in the $R_{\xi}$ Gauge} \label{appendix:propagator}
Here we give the expressions for the propagators in the $R_{\xi}$ gauge.

\noindent
{\bf Scalar Propagator}:
\begin{align}
{\tilde \Delta}^{\hat \phi}_{\{n\}; \{m\}}(k)
= 	\frac{i\delta_{nm}}{-k^{2} - (3\xi - 2)m_{n}^{2}}. \label{eq:propagator01}
\end{align}

\noindent
{\bf Vector Propagator}:
\begin{align}
{\tilde \Delta}^{{\hat h}_{\mu5}}_{\{\mu,n\}; \{\nu,m\}}(k)
&= 	\frac{i\delta_{nm}}{-k^{2} - \xi m_{n}^{2}}
	\left[\eta_{\mu\nu} + \frac{(1-\xi)k_{\mu}k_{\nu}}{-k^{2} - \xi^{2}m_{n}^{2}}\right] \nonumber\\
&= 	{\tilde \Delta}^{({\rm T})}_{\{\mu,n\}; \{\nu,m\}}(k)
	+ {\tilde \Delta}^{({\rm L})}_{\{\mu,n\}; \{\nu,m\}}(k), \label{eq:propagator02}
\end{align}
where
\begin{subequations}
\begin{align}
{\tilde \Delta}^{({\rm T})}_{\{\mu,n\}; \{\nu,m\}}(k)
&= 	\frac{i\delta_{nm}}{-k^{2} - \xi m_{n}^{2}}
	\left[\eta_{\mu\nu} + \frac{k_{\mu}k_{\nu}}{\xi m_{n}^{2}}\right], \label{eq:propagator03a}\\
{\tilde \Delta}^{({\rm L})}_{\{\mu,n\}; \{\nu,m\}}(k)
&= 	\frac{i\xi\delta_{nm}}{-k^{2} - \xi^{2}m_{n}^{2}}
	\left[- \frac{k_{\mu}k_{\nu}}{\xi^{2} m_{n}^{2}}\right]. \label{eq:propagator03b}
\end{align}
\end{subequations}

\begin{widetext}
\noindent
{\bf Graviton Propagator}:
\begin{align}
{\tilde \Delta}^{{\hat h}_{\mu\nu}}_{\{\mu\nu,n\}; \{\rho\sigma, m\}}(k)
&= 	\frac{i\delta_{nm}}{-k^{2} - m_{n}^{2}}
	\biggl[\frac{1}{2}\eta_{\mu\rho}\eta_{\nu\sigma}
	+ \frac{1}{2}\eta_{\mu\sigma}\eta_{\nu\rho}
	- \frac{1}{2}\left(1 - \frac{(1-\xi)m_{n}^{2}}{-k^{2} - (3\xi-2)m_{n}^{2}}\right)
	\eta_{\mu\nu}\eta_{\rho\sigma} \nonumber\\
&	\hspace{5.5em}
	+ \frac{1-\xi}{-k^{2}-\xi m_{n}^{2}}
	\left(\frac{1}{2}\eta_{\mu\rho}k_{\nu}k_{\sigma}
	+ \frac{1}{2}\eta_{\mu\sigma}k_{\nu}k_{\rho}
	+ \frac{1}{2}\eta_{\nu\rho}k_{\mu}k_{\sigma}
	+ \frac{1}{2}\eta_{\nu\sigma}k_{\mu}k_{\rho}\right) \nonumber\\
& 	\hspace{5.5em}
	- \frac{1-\xi}{-k^{2} - (3\xi - 2)m_{n}^{2}}
	(\eta_{\mu\nu}k_{\rho}k_{\sigma}
	+ \eta_{\rho\sigma}k_{\mu}k_{\nu}) \nonumber\\
&	\hspace{5.5em}
	+ \frac{2(1-\xi)^{2}}{-k^{2}-\xi^{2}m_{n}^{2}}
	\left(\frac{1}{-k^{2}-\xi m_{n}^{2}} - \frac{2-\xi}{-k^{2}-(3\xi-2)m_{n}^{2}}\right)
	k_{\mu}k_{\nu}k_{\rho}k_{\sigma}\biggr] \nonumber\\
&= 	{\tilde \Delta}^{({\rm TT})}_{\{\mu\nu,n\}; \{\rho\sigma,m\}}(k)
	+ {\tilde \Delta}^{({\rm T})}_{\{\mu\nu,n\}; \{\rho\sigma,m\}}(k)
	+ {\tilde \Delta}^{({\rm L})}_{\{\mu\nu,n\}; \{\rho\sigma,m\}}(k)
	+ {\tilde \Delta}^{({\rm tr})}_{\{\mu\nu,n\}; \{\rho\sigma,m\}}(k), \label{eq:propagator04}
\end{align}
where
\begin{subequations}
\begin{align}
{\tilde \Delta}^{({\rm TT})}_{\{\mu\nu,n\}; \{\rho\sigma,m\}}(k)
&= 	\frac{i\delta_{nm}}{-k^{2} - m_{n}^{2}}
	\biggl[
	\frac{1}{2}\left(\eta_{\mu\rho} + \frac{k_{\mu}k_{\rho}}{m_{n}^{2}}\right)
	\left(\eta_{\nu\sigma} + \frac{k_{\nu}k_{\sigma}}{m_{n}^{2}}\right)
	+ \frac{1}{2}\left(\eta_{\mu\sigma} + \frac{k_{\mu}k_{\sigma}}{m_{n}^{2}}\right)
	\left(\eta_{\nu\rho} + \frac{k_{\nu}k_{\rho}}{m_{n}^{2}}\right) \nonumber\\
&	\hspace{6em}
	- \frac{1}{3}\left(\eta_{\mu\nu} + \frac{k_{\mu}k_{\nu}}{m_{n}^{2}}\right)
	\left(\eta_{\rho\sigma} + \frac{k_{\rho}k_{\sigma}}{m_{n}^{2}}\right)
	\biggr], \label{eq:propagator05a}\\
{\tilde \Delta}^{({\rm T})}_{\{\mu\nu,n\}; \{\rho\sigma,m\}}(k)
&= 	\frac{(i\xi/2)\delta_{nm}}{-k^{2} - \xi m_{n}^{2}}
	\biggl[
	\left(\eta_{\mu\rho} + \frac{k_{\mu}k_{\rho}}{\xi m_{n}^{2}}\right)
	\left(- \frac{k_{\nu}k_{\sigma}}{\xi m_{n}^{2}}\right)
	+ \left(\eta_{\mu\sigma} + \frac{k_{\mu}k_{\sigma}}{\xi m_{n}^{2}}\right)
	\left(- \frac{k_{\nu}k_{\rho}}{\xi m_{n}^{2}}\right) \nonumber\\
&	\hspace{6em}
	+ \left(\eta_{\nu\rho} + \frac{k_{\nu}k_{\rho}}{\xi m_{n}^{2}}\right)
	\left(- \frac{k_{\mu}k_{\sigma}}{\xi m_{n}^{2}}\right)
	+ \left(\eta_{\nu\sigma} + \frac{k_{\nu}k_{\sigma}}{\xi m_{n}^{2}}\right)
	\left(- \frac{k_{\mu}k_{\rho}}{\xi m_{n}^{2}}\right)
	\biggr], \label{eq:propagator05b}\\
{\tilde \Delta}^{({\rm L})}_{\{\mu\nu,n\}; \{\rho\sigma,m\}}(k)
&= 	\frac{(i\xi^{3}/2)\delta_{nm}}{-k^{2} - \xi^{2}m_{n}^{2}}
	\left[
	4\frac{k_{\mu}k_{\nu}k_{\rho}k_{\sigma}}{\xi^{4}m_{n}^{4}}
	\right], \label{eq:propagator05c}\\
{\tilde \Delta}^{({\rm tr})}_{\{\mu\nu,n\}; \{\rho\sigma,m\}}(k)
&= 	\frac{(i/6)\delta_{nm}}{-k^{2} - (3\xi - 2)m_{n}^{2}}
	\left[- \left(\eta_{\mu\nu} - 2\frac{k_{\mu}k_{\nu}}{m_{n}^{2}}\right)
	\left(\eta_{\rho\sigma} - 2\frac{k_{\rho}k_{\sigma}}{m_{n}^{2}}\right)\right]. \label{eq:propagator05d}
\end{align}
\end{subequations}
\end{widetext}
\noindent
Remarks:
\begin{enumerate}
\item These propagators are calculated with the canonically normalized fields, $h_{\mu\nu} \mapsto {\hat h}_{\mu\nu} = (M_{\rm Pl}/\sqrt{2})h_{\mu\nu}$, $h_{\mu5} \mapsto {\hat h}_{\mu5} = M_{\rm Pl}h_{\mu5}$ and $\phi \mapsto {\hat \phi} = (\sqrt{3}M_{\rm Pl}/2)\phi$.

\item By taking the limit $\xi\to\infty$ these propagators just reduces to those calculated in the unitary gauge.

\item The unitary gauge limit $\xi\to\infty$ and the massless limit $m_{n}\to0$ do not commute with each other, just as in ordinary spontaneously broken gauge theories.

\item $\Delta^{({\rm TT})}$ is nothing but the propagator for the 4d massive graviton with the Fierz-Pauli mass term.
\end{enumerate}
%%%%% REFERENCES %%%%%

\end{document}